\documentclass[10pt,conference,compsocconf,letterpaper]{IEEEtran}
\makeatletter
\def\ps@headings{%
\def\@oddhead{\mbox{}\scriptsize\rightmark \hfil \thepage}%
\def\@evenhead{\scriptsize\thepage \hfil \leftmark\mbox{}}%
\def\@oddfoot{}%
\def\@evenfoot{}}
\makeatother 

\usepackage{cite}
\usepackage{times}
\usepackage{dsfont}
\usepackage{cite}
\usepackage{times}
\usepackage{amsthm}
\usepackage{graphicx}
\usepackage{subfigure}
\usepackage{color}
\usepackage{ifpdf}
\usepackage{CJK}
\usepackage{epsfig}
\usepackage{latexsym}
\usepackage{amsfonts}
\usepackage{amssymb}
\usepackage{paralist}
\usepackage{comment}
\usepackage{xspace}
\usepackage{mathrsfs}
\usepackage{amssymb}
\usepackage{diagbox}
\usepackage{setspace}
\usepackage{color}
\usepackage[small]{caption2}
\usepackage{balance}
\usepackage{algorithm}
\usepackage[noend]{algpseudocode}

\usepackage{amssymb}
\usepackage{url,epsfig,array}
\usepackage{leftidx}
\usepackage{amsmath}
\usepackage{url}
\usepackage{amsmath}
\usepackage[
top=0.65in, bottom=0.65in,
left=0.65in, right=0.65in
]{geometry}
\def\ie{\textit{i.e.}\xspace}
\def\etal{\textit{et al.}\xspace}

\def\eg{\textit{e.g.}\xspace}

\def\vs{\textit{vs.}\xspace}
\def\wrt{\textit{w.r.t.}\xspace}

\newtheorem{corollary}{Corollary}

\newtheorem{lemma}{Lemma}
\newtheorem{remark}{Remark}

\newtheorem{assumption}{Assumption}
\begin{document}

%\begin{spacing}{1.01}
%
% paper title
% can use linebreaks \\ within to get better formatting as desired

\title{Adaptive Configuration for Heterogeneous Participants in Decentralized Federated Learning}

\author{\IEEEauthorblockN{Yunming Liao$^{1,2}$ \ \ \ $^*$Yang Xu$^{1,2}$ \  \ \  Hongli Xu$^{1,2}$ \ \ \ Lun Wang$^{1,2}$ \ \  \   Chen Qian$^3$}
\IEEEauthorblockA{
Email:\{ymliao98, wanglun0\}@mail.ustc.edu.cn, \{xuyangcs, xuhongli\}@ustc.edu.cn, cqian12@ucsc.edu\\
$^1$School of Computer Science and Technology, University of Science and Technology of China, China\\
$^2$Suzhou Institute for Advanced Study, University of Science and Technology of China, China\\
$^3$Department of Computer Science and Engineering, Jack Baskin School of Engineering, University of California, Santa Cru.
} }

\maketitle

\begin{abstract}
Data generated at the network edge can be processed locally by leveraging the paradigm of edge computing (EC).
Aided by EC, decentralized federated learning (DFL), which overcomes the single-point-of-failure problem in the parameter server (PS) based federated learning, is becoming a practical and popular approach for machine learning over distributed data.
However, DFL faces two critical challenges, \ie, system heterogeneity and statistical heterogeneity introduced by edge devices.
To ensure fast convergence with the existence of slow edge devices, we present an efficient DFL method, termed FedHP, which integrates adaptive control of both local updating frequency and network topology to better support the heterogeneous participants.
We establish a theoretical relationship between local updating frequency and network topology regarding model training performance and obtain a convergence upper bound.
Upon this, we propose an optimization algorithm, that adaptively determines local updating frequencies and constructs the network topology, so as to speed up convergence and improve the model accuracy.
Evaluation results show that the proposed FedHP can reduce the completion time by about 51\% and improve model accuracy by at least 5\% in heterogeneous scenarios, compared with the baselines.

\end{abstract}
% \vspace{-0.1em}
% \vspace{-0.1em}

\IEEEpeerreviewmaketitle

% \vspace{-0.6em}
\section{Introduction}\label{sec:intro}
The past few years have witnessed remarkable advancements in mobile computing and the Internet of Things.
Mobile devices constantly generate massive data, such as photos and voices, which are of great value for developing intelligent applications \cite{lim2020federated,wang2020optimizing}.
Meanwhile, edge computing (EC) systems have been deployed to store data locally and push more computing power to the network edge for data analysis \cite{shi2016edge, satyanarayanan2017emergence, 9484767}.
With the emergence of EC, federated learning (FL) \cite{kairouz2019advances,mcmahan2017communication,park2019wireless,yang2019federated} has been developed to perform distributed model training at the network edge or end devices close to the data source.
FL does not only prevent personal privacy from being exposed but also fully utilizes plenty of computation resources at the network edge.

Traditional FL requires a parameter server (PS) to communicate with the edge nodes (\ie, participants) \cite{mcmahan2017communication,konevcny2016federated, wang2019adaptive}, and involves model transmission from a certain (possibly large) number of nodes for model aggregation, which brings enormous amount of traffic workload to the PS.
Consequently, the PS may become the system bottleneck, leading to the risk of network congestion and poor scalability.
In comparison, decentralized federated learning (DFL) \cite{kairouz2019advances,lian2017can,yu2019parallel,hua2022overlay} is becoming an attractive solution by disseminating information through peer-to-peer (P2P) communication, to avoid the communication bottleneck at the centralized server.
Moreover, since there is no need to forward the local models from nodes to the PS, the potential of single point failure can be avoided and the system scalability will be significantly improved.
This work focuses on DFL and explores its communication and computation efficient learning strategies so as to enhance model training at the network edge.

There are two important features in EC systems making it difficult to implement efficient DFL.
1) \textit{System Heterogeneity.} In EC, the capabilities of edge nodes are usually limited and heterogeneous \cite{lim2020federated,zhang2018adaptive}.
There could be a tenfold difference in computing capabilities (\eg, CPU frequency) or communication capabilities (\eg, bandwidth, throughput) among edge nodes \cite{ma2021adaptive, ma2021fedsa,xu2022adaptive}.
Due to system heterogeneity, fast edge nodes may have to wait for the stragglers in a synchronous manner, which incurs non-negligible waiting time and deteriorates training efficiency.
2) \textit{Statistical Heterogeneity.} The local data collected by edge nodes usually depends on their functions and/or locations, resulting in non- independent and identically distributed (non-IID) local data across all edge nodes.
The non-IID data (known as statistical heterogeneity) will decelerate the convergence rate and even compromise the accuracy of trained models \cite{zhao2018federated, wang2020optimizing, wang2022accelerating}.

In general, edge nodes always update the models with their globally-synchronized neighbor models, which is proven to achieve similar convergence rate (\wrt the number of rounds/iterations) as the parallel mini-batch SGD, and will converge to satisfied solutions with high test accuracy \cite{yu2019parallel}.
Besides, given limited capabilities on edge nodes, a synchronous DFL method, named LD-SGD \cite{li2019communication}, has been proposed, which alternates the frequencies of local updating and global updating to significantly reduce the communication resource consumption.
As for statistical heterogeneity, Onoszko \etal \cite{onoszko2021decentralized} proposed a synchronous method named performance-based neighbor selection (PENS), where nodes with similar data distributions communicate with each other.
However, the synchronization barrier of these methods often leads to idle time for staying and waiting for the stragglers (\ie, the slow participants) before model aggregation, especially in the heterogeneous system.
Moreover, PENS always suffers from more computing time for neighbor selection (\ie, network topology construction) and model training at each communication round due to system heterogeneity.
Although the asynchronous DFL \cite{lian2018asynchronous, assran2020asynchronous,luo2020prague, zhou2021communication} contributes to addressing the challenge of system heterogeneity and accelerating the convergence rate \wrt time, each node receives and aggregates the stale models, which amplifies the negative impact of non-IID data on test accuracy and even leads to model divergence \cite{ma2021fedsa}.
Herein, we focus on the synchronous implementation of DFL to cope with the potential problems, such as delayed convergence time and compromised model accuracy, caused by system and statistical heterogeneities.

In this paper, we investigate the benefits of controlling local updating frequency and network topology, which are jointly optimized to adequately address the two heterogeneity issues for synchronous DFL.
Unlike the identical local updating frequency and fixed neighbors (\ie, network topology) for all edge nodes \cite{lian2017can,li2019communication}, we explore to adaptively assign different local updating frequencies for heterogeneous nodes and adjust network topology to eliminate the idle time incurred by synchronization.
The coupled relationship between local updating frequency and network topology will be elaborated in Sec. \ref{sec:relation}.
According to our theoretical analysis and pretest in Sec. \ref{sec:analysis}, a relatively smaller or larger local updating frequency will lead to more communication rounds or lower model accuracy.
Therefore, as training progresses, it is necessary yet challenging to simultaneously determine the appropriate local updating frequencies and neighbors for different edge nodes so as to well balance the trade-off between convergence rate and model accuracy.
The main contributions of this paper are summarized as follows:
% \vspace{-0.075cm}
\begin{itemize}
    \item We design an efficient DFL method, called  FedHP, which integrates adaptive control of local updating frequency and network topology to better overcome the challenges of system and statistical heterogeneities in EC systems.
    \item We theoretically analyze the convergence rate and obtain a convergence upper bound related to local updating frequency and network topology.
    % We establish a theoretical relationship between local updating frequency and network topology in terms of model training performance, and obtain a convergence upper bound.
    Upon this, we propose a control algorithm, which adaptively determines appropriate local updating frequencies and neighbors for different edge nodes, so as to speed up training and improve the model accuracy.
    \item The performance of our method is evaluated through extensive simulation experiments.
    The evaluation results show that our method can reduce the convergence time by about 51\% and improve model accuracy by at least 5\% in heterogeneous scenarios, compared to existing DFL methods.
\end{itemize}

The rest of this paper is organized as follows.
Sec. \ref{sec:prelim} formalizes the optimization problem in FedHP.
Sec. \ref{sec:analysis} gives the convergence analysis of FedHP.
Based on the analysis, we propose an efficient algorithm in Sec. \ref{sec:alg}.
Then in Sec. \ref{sec:results}, we report our experimental results.
We discuss some related works in Sec. \ref{sec:related} and conclude the paper in Sec. \ref{sec:conclusion}.

% \vspace{-0.7em}
\section{Preliminaries and Problem Formulation}\label{sec:prelim}
% \vspace{-0.3em}
\subsection{Network Model}
% \vspace{-0.3em}
% \begin{table}[t]
%     \setlength{\abovecaptionskip}{10pt}%
%     \setlength{\belowcaptionskip}{0pt}%
%     \centering
%     \caption{Key Notations.}
%     \label{tbl:notation}
%     \begin{tabular}{cl}
%     \hline
%         \textbf{Symbol} & \textbf{Semantics}\\
%     \hline
%         $\mathcal{V}$ & the set of workers\\
%         $H$ & total number of communication rounds\\
%         $E^{h}$ & the set of links at round $h$, and $E^{h} \subset \mathcal{V} \times \mathcal{V}$\\
%         $a_{i,j}^{h}$ & whether link $e_{i,j} \in E^{h}$ or not\\
%         $\mathcal{N}^{h}_i$ & neighbor set of worker $i$ at round $h$\\
%         $x_i^{h}$ & local model of worker $i$ at round $h$\\
%         %$h$ & the global iteration step}\\
%         $t^h$ & completion time of round $h$\\
%         $t_i^h$ & completion time of round $h$ on worker $i$\\
%         $\mathcal{W}^h$ & average waiting time at round $h$\\
%         $\mu_i^h$ & computing time of one local iteration of\\ &worker $i$ at round $h$\\
%         $\eta_{i,j}^h$ & communication time between worker $i$ \\ & and worker $j$ at round $h$\\
%         $\tau_i^h$ & local updating frequency of worker $i$ \\ & at round $h$\\
%         $\varepsilon$ & waiting time threshold\\
%         $D_{i,j}^{h}$ & consensus distance between two local models \\ &  of worker $i$ and worker $j$ at round $h$\\		
% 		$D^{h}_i$ & consensus distance between the local model \\ &  of worker $i$ and the average model at round $h$\\
%     \hline
%     \end{tabular}
% %    \vspace{-0.5cm}
% \end{table}
An EC system includes a set of distributed workers (\eg, IoT devices or small base stations) $\mathcal{V} = \{v_1,v_2,\ldots,v_N\}$, with $|\mathcal{V}|=N>1$.
% $\mathbb{D}$
In DFL, the workers collaboratively train deep learning models on their local datasets, and each worker needs to exchange models with its neighbors rather than sharing its original data.
% , and some techniques, \eg, secure aggregation \cite{bonawitz2017practical} and differential privacy \cite{abadi2016deep}, can be adopted to preserve the privacy of transmitted models or gradients better.
A control node (\ie, coordinator) is still needed to collect the global information about model training statuses and network conditions in DFL \cite{zhou2021communication,wang2019matcha,wang2022accelerating, xu2021decentralized}. However, unlike the parameter server in FL, the coordinator does not aggregate the models and hence will not become the bandwidth bottleneck.
Furthermore, any worker can act as the coordinator.
%  to make decisions on local updating frequencies and topology construction.
%It should be noted that the coordinator is quite different from the PS needing to aggregate local gradients and update the global model.}
% The coordinator (\eg, cloud server) only collects the information about model training and network conditions to make decisions on local updating frequencies for workers and topology construction.
% The required information at the logical coordinator includes the link speeds among workers, consensus distances among workers and the time about local training of each worker.
Since the size of these information (\eg, 100-300KB \cite{lyu2018multi}) is much smaller than that of model parameters, it is reasonable to ignore the cost (\eg, bandwidth consumption and time cost) for information collection \cite{lyu2019optimal}.
% the number of available channel subcarriers in a network, the bandwidth resource at each edge node, and the basic system topology.
% For ease of expression, we list some important notations in Table \ref{tbl:notation}.
%, and $h$ in each notation denotes the index of training epoch without any confusion.

The P2P network topology at the $h$-th communication round can be expressed as a connected undirected graph $\mathcal{G}^h=(\mathcal{V},E^h)$, where $\mathcal{V}$ denotes the worker set and $E^{h}$ denotes the set of links connecting workers at communication round $h$.
Specifically, the P2P network topology at round $h$ can be expressed as a symmetric adjacency matrix $\mathbf{A}^{h} = \{a_{i,j}^{h} \in \{0, 1\}, 1 \leq i,j \leq N\}$, where $a_{i,j}^{h} = 1$ if $e_{i,j}^{h} \in E^{h}$, otherwise $0$.
The neighbor set of worker $i$ at round $h$ is represented as $\mathcal{N}^{h}_i$, whose cardinality is denoted as $|\mathcal{N}^{h}_i|=\sum_{j\in \mathcal{N}^{h}_i} a_{i,j}^{h}$.
The degree matrix $\mathbf{D}^{h}=\{d^{h}_{i, j}, 1 \leq i,j \leq N\}$ is defined as a diagonal matrix, where $d^{h}_{i, i} = |\mathcal{N}^{h}_i|$.
Combining the adjacency matrix and the degree matrix, the Laplacian matrix $\mathbf{L}^{h}$ can be expressed as follows:
\begin{equation}
    \mathbf{L}^{h} = \mathbf{D}^{h} - \mathbf{A}^{h}.
\end{equation}
%Then we express Laplacian matrix as $\mathbf{L}^{h} = \mathbf{D}^{h} - \mathbf{A}^{h}$.
According to the spectral graph theory \cite{chung1997spectral},  $\lambda_2(\mathbf{L}^{h}) > 0$ if and only if the topology is connected, where $\lambda_{m}(\mathbf{L}^{h})$ denotes the $m$-th smallest eigenvalue of matrix $\mathbf{L}^{h}$.

% \vspace{-0.6em}
\subsection{Model Training Process}
% \vspace{-0.2em}
In DFL, worker $i$ updates the local model parameter $x_i$ at the $h$-th communication round based on a mini-batch $\xi_i$ sampled from its local dataset $\mathcal{D}_i$.
% Let $f_i(x_i)$ and $F_i(x_i)$ (\ie, $F_i(x_i;\xi_i)$) denote the local loss function and the loss function over mini-batch $\xi_i$, respectively.
Let $f_i(x_i)$ and $F_i(x_i;\xi_i)$ (for ease of description, written as $F_i(x_i)$) denote the local loss function and the loss function over mini-batch $\xi_i$, respectively.
Generally, model training can be formally described as optimizing the following objective function \cite{koloskova2019decentralized}:
\begin{equation}\label{Eq:loss function}
f^* := \min_{x \in \mathbb{R}^d}\ [\ f(x) := \frac{1}{N} \sum_{i=1}^{N} f_i(x_i)\ ]\mbox{,}
\end{equation}
where $f_i(x_i) := \mathbb{E}_{\xi_i \sim \mathcal{D}_i}\ F_i(x_i)$ and $x$ denotes the global model parameter.
This setting covers the important cases of empirical risk minimization in DFL \cite{koloskova2019decentralized}.

The model will be updated by applying the decentralized stochastic gradient descent (DSGD) algorithm \cite{tsitsiklis1986distributed}, which provides an effective way to optimize the loss function in a decentralized manner.
For the mini-batch stochastic gradient descent, a gradient descent step over a mini-batch on each worker is regarded as a local iteration (or a local update).
After performing one or multiple local iterations, each worker exchanges local models or gradients with its neighbors and aggregates these models.
Such a training process is regarded as a communication round.
$x_i^{h,k}$ denotes the local model of worker $i$ at the $k$-th local iteration within communication round $h$.
At the beginning of communication round $h$, by setting $x_i^{h, 0}=x_i^{h}$, worker $i$ updates its local model by gradient descent as follows \cite{wang2022accelerating,xu2022adaptive}:
\begin{equation} \label{Eq:Update rule 2}
x_i^{h, k+1} = x_i^{h, k} - \eta \nabla F_i(x_i^{h, k})\mbox{,} \ 0 \le k < \tau \mbox{,}
\end{equation}
where $\eta$ is the local learning rate, $\tau$ is the local updating frequency, and $\nabla F_i(x_i^{h, k})$ is the gradient. 
The local updates of worker $i$ at round $h$ is denoted as $g_i^{h} = \sum_{k=0}^{\tau-1}\nabla F_i(x_i^{h, k})$.
Then the local updating of worker $i$ can be rewritten as:
\begin{equation} \label{Eq:Update rule}
    x_i^{h+1} = x_i^{h} - \eta \cdot g_i^h.
\end{equation}

After local updating, workers send local models to their neighbors.
Based on the received model parameters, worker $i$ will aggregate these models from neighbors:
%  in an element-wise way
\begin{equation} \label{Eq:Update rule 1}
	x_i^{h+1} = x_i^{h} + \sum_{j \in \mathcal{N}_i^{h}} w^{h}_{i,j} (x_j^{h} - x_{i}^{h})\mbox{,}
\end{equation}
where $\mathcal{N}_i^{h}$ is the neighbor set of worker $i$ at round $h$ and $w_{i,j}^{h}, j \in \mathcal{N}_i^{h}$, is the mixing weight for aggregating the model of neighbor $j$.
Defining $u_{max}^{h}$ as the maximum of $|\mathcal{N}_i^{h}|$ over workers at round $h$, a simple suboptimal choice of $w^{h}_{i,j}$ is \cite{xiao2004fast}:
\begin{equation} \label{Eq:Step size}
	w^{h}_{i,j} = \frac{1}{u_{max}^{h} + 1}.
\end{equation}

\subsection{Consensus Distance}
Unlike the traditional PS architecture, there is no global model in DFL, and local models hosted by different workers are not always the same.
We introduce the \emph{consensus distance} metric to measure the discrepancy among local models \cite{koloskova2019decentralized,lin2021on,wang2022accelerating}.
%  and the average of all local models 
%how far a local model deviates from the average of all local models \cite{koloskova2019decentralized,lin2021on}.
%the average discrepancy between each node and the mean of model parameters over all machines.
Firstly, the consensus distance between model of worker $i$ and model of worker $j$ at the $h$-th communication round is defined as:
\begin{equation} \label{Eq:Consenus Distance}
    D^{h}_{i,j} = \left \| x_i^h - x_j^h \right \|.
\end{equation}
Then the consensus distance between local model of worker $i$ and ``global model'' (\ie, the average of all workers' models) at round $h$ is defined as:
\begin{equation} \label{Eq:Local Consenus Distance}
D^{h}_i = \left \| \overline{x}^{h} - x_i^{h} \right \|\mbox{,}
\end{equation}
where $\overline{x}^{h} = \frac{1}{N} \sum_{i=1}^{N} x_i^{h}$ denotes the average of all workers' models at round $h$.
It is worth noting that $\overline{x}^{h}$ is not available in practice because there is no PS to collect all workers' models in DFL. 
To this end, we would estimate $D_i^{h}$ using consensus distance between the local model of worker $i$ and the models of its neighbors (\ie, $D_{i,j}^{h}, j \in \mathcal{N}_i^h$), which will be elaborated in Sec. \ref{subsec_distance_estimation}.
% More details are explained in Section \ref{subsec:upper bound of dist}.
Accordingly, the average consensus distance of all workers' models is:
\begin{equation} \label{Eq:Avg Consenus Distance}
D^{h} = \frac{1}{N} \sum_{i=1}^{N} D^{h}_i.
\end{equation}
Similar to the weight divergence \cite{zhao2018federated, qian2020towards} in the PS architectures, the consensus distance is correlated to data distribution and is the key factor that captures the joint effect of decentralization \cite{lin2021on}, which motivates us to apply consensus distance for topology construction to overcome the challenge introduced by non-IID data.

% Under the decentralized FL setting, workers exchange models with neighbors to reach consensus among all workers \cite{london2019logarithmic}.
% Since the consensus distance plays a key role in decentralized FL \cite{lin2021on}, it is a natural strategy to improve training performance by exchanging models between two workers with significantly different data distributions \cite{hsieh2020non}, \ie, with large consensus distance.

\subsection{Relationship between Local Updating Frequency and Network Topology}\label{sec:relation}
% In this section, we explain the coupling relationship between the local updating frequency and network topology. 
% Firstly, the local models trained with different local updating frequencies are discrepant, which requires to select matched neighbors for model aggregation to achieve satisfied model accuracy.
% Secondly, the completion time of communication round (include computing time and communication time) varies with dynamic network topology, which requires to assign suitable local updating frequencies for heterogeneous workers to reduce the waiting time. 
% Accordingly, we propose to jointly optimize the local updating frequency and network topology to address the system heterogeneity and statistical heterogeneity in DFL.
In this section, we explain the coupled relationship between local updating frequencies and network topologies. 
On the one hand, the computing time of one local iteration and the transmission time of one model among workers are highly different due to system heterogeneity. 
% As illustrated in Section \ref{subsec_system_test}, t
% The highest-performance (or fastest) worker can be 10 times faster than the lowest-performance (or slowest) one.
However, in traditional synchronous schemes, local updating frequencies among workers are usually identical or fixed at each communication round.
Accordingly, fast workers have to wait for slow ones, incurring non-negligible idle time and significantly reducing the training efficiency \cite{ma2021adaptive, zhang2018adaptive}.
Considering the heterogeneous computing capabilities of workers, before aggregation, the workers with higher computing capabilities will perform more local iterations while the workers with lower computing capabilities only perform fewer local iterations.
%which is usually restricted by synchronization barrier.
On the other hand, data samples across all workers may be non-IID, which seriously affects the convergence rate and even compromises the accuracy of trained model \cite{zhao2018federated, wang2020optimizing}.
To deal with the statistical heterogeneity, the workers with significantly different data distributions (\ie, with large consensus distance) can be connected preferentially and frequently.
After that, the training performance over non-IID data can be guaranteed meanwhile the waiting time and training time among workers would be significantly reduced.
% posing additional challenges to the convergence and speed of decentralized FL.
%  non-IID data would significantly reduce convergence accuracy \cite{zhao2018federated}.

Furthermore, the local models trained with different local updating frequencies are discrepant, which requires to select suitable neighbors for model aggregation to achieve satisfied model accuracy.
Meanwhile, the completion time of each communication round (including computing time and communication time) varies with dynamic network topology, which requires to assign appropriate local updating frequencies for heterogeneous workers to reduce the waiting time. 
Accordingly, we propose to jointly optimize the local updating frequency and network topology to address the system heterogeneity and statistical heterogeneity in DFL.
% Accordingly, we propose to dynamically adjust local updating frequencies for different workers and network topologies at each communication round, so as to address the system heterogeneity and statistical heterogeneity in DFL.

% Under the DFL setting, workers exchange models with neighbors to reach consensus among all workers \cite{london2019logarithmic}.
% Since the consensus distance plays a key role in decentralized FL \cite{lin2021on}, it is a natural strategy to improve training performance by exchanging models between two workers with significantly different data distributions \cite{hsieh2020non}, \ie, with large consensus distance.
% Considering the heterogeneous computing capabilities of workers, before aggregation, the workers with higher computing capabilities can perform more local iterations while the workers with lower computing capabilities only perform less local iterations.
% % During model transmission, the workers communicates with neighbors by links with higher communication capabilities.
% To deal with the statistical heterogeneity, the workers with significantly different data distributions (\ie, with large consensus distance) can be connected preferentially and frequently.
% After that, the training performance of non-IID data can be guaranteed meanwhile the waiting time and traning time among workers would be significantly reduced.

\subsection{Problem Formulation}
% This section defines the problem of DFL with adaptive local updating and network topology given the challenges of system heterogeneity and statistical heterogeneity.
This section defines the problem of efficient DFL with adaptive local updating and network topology: \textit{minimizing the training time while requiring workers to achieve a satisfied accuracy for their models}. 
Given a DFL task in the EC system, we need to determine the local updating frequencies and average consensus
distance of all workers to minimize the training time.
%  $\sum_{h=1}^{H} t^h$
First, the local updating frequency and the computing time of one local iteration at the $h$-th communication round on worker $i$ are denoted as $\tau_i^h$ and $\mu_i^h$, respectively.
Let $\mathbf{B}^{h} = \{\beta_{i,j}^{h}, 1 \leq i,j \leq N\}$ denote the communicating time matrix at round $h$, where $\beta_{i,j}^h$ is the communicating time between worker $i$ and worker $j$.
% =\beta_{j,i}^h
% $\mathbf{\beta}^h$ denotes the communicating time matrix at round $h$, and $\beta_{i,j}^h$ ($\beta_{i,j}^h=\beta_{j,i}^h$) denotes the communicating time between worker $i$ and worker $j$.
Therefore, the local updating time (including computing time and communication time) of worker $i$ at round $h$ is formulated as:
\begin{equation}
    t_i^h=\tau_i^h \cdot \mu_i^h + \max\{\beta_{i,j}^h\}\ \forall i \in [N], \forall j \in \mathcal{N}^{h}_i.
\end{equation}
In addition, the waiting time of worker $i$ can be expressed as $t^h-t_i^h$, where $t^h=\max\{t_i^h\}\  (\forall i \in [N])$ denotes the local updating time of the slowest worker at round $h$.
$t^h$ also denotes the completion time of round $h$.
Then the average waiting time of all workers at round $h$ can be formulated as:
\begin{equation}
    \mathcal{W}^h =\frac{1}{N} \sum_{i=1}^{N}(t^h-t_i^h).
\end{equation}
% Our method ensures that the average waiting time will be small enough to mitigate the effects of the synchronization barrier.
% Considering the heterogeneous computing capabilities of workers, before aggregation, the workers with higher computing capabilities can perform more local iterations while the workers with lower computing capabilities only perform less local iterations, so that the average waiting time will be small enough to mitigate the effects of the synchronization barrier.
% To deal with the statistical heterogeneity, the workers with significantly different data distributions (\ie, with large consensus distance) can be connected preferentially and frequently.
% After that, the training efficiency can be improved, and the training performance of non-IID data can be guaranteed.
Accordingly, we formulate the problem as follows:

\vspace{0.1cm}
\centerline{$\min \sum\limits_{h=1}^{H} t^h$}
\vspace{-0.4cm}
\begin{equation}\label{problem}
    s.t.
    \begin{cases}
    D^{h+1} \le D_{max}^{h}, \\
    \lambda_2(\mathbf{L}^{h}) > 0,\\
    t_i^h=\tau_i^h \cdot \mu_i^h+ \max\{\beta_{i,j}^h\}, \forall i \in [N], \forall j \in \mathcal{N}^{h}_i\\ 
    \mathcal{W}^h =\frac{1}{N} \sum_{i=1}^{N}(t^h-t_i^h) \le \varepsilon
    % , \quad\quad\ \  \forall h \in  [H]\\
    \end{cases}
\end{equation}
% where $\sum\limits_{h=1}^{H} t^h$ denotes the model training time.
The first inequality expresses that the average consensus distance should not exceed the predefined threshold $D_{max}^{h}$.
We set $D_{max}^{h}$ as the same in \cite{lin2021on} and the details are described in Sec. \ref{sec:alg}.
The second inequality ensures a connected topology in each communication round, which is essential to guarantee the training convergence \cite{pmlr-v119-koloskova20a}.
The third set of equalities denotes the formulation of the local updating completion time and communication time on worker $i$ at the $h$-th communication round, where $\beta_{i,j}$ denotes the communication time between worker $i$ and worker $j$.
The fourth set of inequalities essentially guarantees that the average waiting time of all workers at each communication round is sufficiently small, where $\varepsilon > 0$ is the time threshold, so as to mitigate the effects of the synchronization barrier. Our objective is to minimize the training time under the constraints.

% \vspace{-0.2em}
\section{Convergence Analysis}\label{sec:analysis}
In this section, we analyze the model convergence rate of our method in theory and obtain a convergence upper bound related to local updating frequency and network topology.
We first make the following assumptions, which are widely used in previous works \cite{tang2018communication, tang2019texttt, pmlr-v119-koloskova20a, wang2022accelerating}:

\begin{assumption}\label{Assumption 1} (\textbf{\textit{L-smooth}}) Each local objective function $f_i : \mathbb{R}^d \rightarrow \mathbb{R}$ on workers is $L$-smooth:
\begin{equation}
    \left\| \nabla f_i(y) - \nabla f_i(x)\right\|_2 \leq L \left\| y - x \right\|_2, \forall x,y \in \mathbb{R}^d.
\end{equation}
\end{assumption}

\begin{assumption}\label{gradient}
    (\textbf{\textit{Unbiased Local Gradient Estimator}}) Let $\xi_i^h$  be a random local data sample at the
    $h$-th communication round on worker $i$. The local gradient estimator is unbiased as follows:
    \begin{equation}
        \mathbb{E}\left[\nabla F_{i}\left(x_i^{h}, \xi_{i}^{h}\right)\right]=\nabla f_{i}\left(x_i^{h}\right). %,
    \end{equation}
\end{assumption}

\begin{assumption}\label{Assumption 2} (\textbf{\textit{Bounded gradient variance}}) The variance of stochastic gradients at each worker is bounded:
\begin{align}
    \mathbb{E}\left\|\nabla F_i(x_i, \xi_i) - \nabla f_i(x_i)\right\|_2^2 \leq \sigma^2, \forall x \in \mathbb{R}^d, \forall i \in [N]\mbox{,} \label{assump2-1}\\
    \frac{1}{N} \sum_{i=1}^{N} \left\|\nabla f_i(x_i) - \nabla f(x)\right\|_2^2 \leq \zeta^2, \forall x \in \mathbb{R}^d, \forall i \in [N]. \label{assump2-2}
\end{align}
\end{assumption}
The variance in Eq. \eqref{assump2-1} denotes how far the estimated gradient over mini-batch $\xi_i$ deviates from the true gradient of $f_i(x_i)$. 
In addition, $\zeta$ in Eq. \eqref{assump2-2} indicates the degree of difference between local functions on workers and the global function $f(x)$, indicating the heterogeneity of the non-IID datasets among different workers.
%For ease of analysis, we denote $\zeta = \sum_{i=1}^m \frac{\zeta_i}{m}$. 
In particular, if the data distributions across workers are IID, all functions are identical (\ie, $f_i(x_i) = f_j(x_j), \forall i,j \in [N]$), thus $\zeta=0$.

% We use a universal bound $\zeta$ to quantify the heterogeneity of the non-IID datasets among different workers.
% In particular, $\zeta=0$ corresponds to IID datasets.

\begin{assumption}\label{Assumption 3} (\textbf{\textit{Spectral gap}}) The weight matrix $W$ is symmetric doubly stochastic. We define $\rho = \max \{|\lambda_2(W)|$, $ |\lambda_N(W)| \}$ and assume $\rho < 1$.
\end{assumption}

\begin{lemma}
Under the above assumptions with $\eta \le \frac{1}{4L \tau}$, we have the following expression:
\begin{align}\label{eq:lemma1}
\mathbb{E}f(\overline{x}^{h+1}) & \le f(\overline{x}^{h})  - \frac{\eta \tau}{4} \left \| \nabla f(\overline{x}^{h}) \right \|_2^2 \notag \\
& + \frac{\eta L^2 \tau}{N}  \sum_{i=1}^{N} \left \| \overline{x}^{h} - x_i^{h} \right \|_2^2 + \frac{\sigma^2 \eta^2  \tau^2 L}{N}\mbox{,} %\frac{\sigma^2 }{16 L N}
\end{align}
where $\tau=\max\{\tau_i^h\}$.
\end{lemma}

\begin{IEEEproof}
For convenience, we introduce the following matrix notations:
\begin{equation}
	\left\{
	\begin{array}{ll}
		X^{h} := [x_1^{h}, \dots, x_N^{h} ] , \\
		\overline{X}^{h} := [\overline{x}^{h}, \dots, \overline{x}^{h} ], \\
		\nabla F(X^{h}) := [\nabla F_1(x_1^{h}), \dots,  \nabla F_N(x_N^{h})],
	\end{array}
	\right.
\end{equation}
where $\overline{x}^{h} = \frac{1}{N} \sum_{i=1}^{N} x_i^{h}$ and $\overline{\nabla F}(X^{h}) = \frac{1}{N} \sum_{i=1}^{N}\nabla F_i(x_i^{h}) $.

% Obviously, averaging with the mixing matrix $W$  preserves the average of models across workers, \ie, 
% \begin{equation} \label{eq:averaging}
% XW \frac{\mathbf{1}\mathbf{1}^T}{N} = X \frac{\mathbf{1}\mathbf{1}^T}{N}
% \end{equation}
% Following Eq. \eqref{eq:averaging} and Assumption 1, we obtain:
According to the Lipschitz smoothness property in Assumption 1, we obtain:
\begin{align} \label{eq:proof-1-1}
\mathbb{E}f(\overline{x}^{(h+1)}) &= \mathbb{E}f \left(\overline{x}^{h} - \frac{\eta \tau}{N}\sum_{i=1}^{N}  \nabla F_i(x_i^{h}) \right) \notag \\
&\le f(\overline{x}^{h}) - \tau \mathbb{E} \left\langle \nabla f(\overline{x}^{h}), \frac{\eta}{N}\sum_{i=1}^{N}  \nabla F_i(x_i^{h}) \right\rangle \notag \\
&+ \frac{L\eta^2 \tau^2}{2} \mathbb{E} \left \| \frac{1}{N}\sum_{i=1}^{N}  \nabla F_i(x_i^{h}) \right \|_2^2.
\end{align}

Then we bound the second term:
\begin{align} \label{eq:proof-1-2}
&- \mathbb{E} \langle \nabla f(\overline{x}^{h}), \frac{\eta}{N}\sum_{i=1}^{N}  \nabla F_i(x_i^{h}) \rangle \notag \\
&= \mathbb{E} \langle \nabla f(\overline{x}^{h}), \eta  \nabla f(\overline{x}^{h})-\frac{\eta}{N}\sum_{i=1}^{N} \nabla F_i(x_i^{h}) - \eta \nabla f(\overline{x}^{h})\rangle \notag \\
&= \mathbb{E} \langle \nabla f(\overline{x}^{h}), \frac{\eta}{N}\sum_{i=1}^{N} \nabla f_i(\overline{x}^{h})-\frac{\eta}{N}\sum_{i=1}^{N}  \nabla F_i(x_i^{h}) - \eta \nabla f(\overline{x}^{h})\rangle \notag \\
&= \langle \nabla f(\overline{x}^{h}),\frac{\eta}{N} \sum_{i=1}^{N} \left(\nabla f_i(\overline{x}^{h}) - \nabla f_i(x_i^{h})\right) \rangle - \eta  \left \| \nabla f(\overline{x}^{h}) \right \|^2_2 \notag \\
&= \frac{\eta}{N} \sum_{i=1}^{N}  \left \langle \nabla f(\overline{x}^{h}), \nabla f_i(\overline{x}^{h}) - \nabla f_i(x_i^{h}) \right \rangle - \eta  \left \| \nabla f(\overline{x}^{h}) \right \|^2_2 \notag \\
&\le \frac{\eta}{2 N} \sum_{i=1}^{N} \left \| \nabla f_i(\overline{x}^{h}) - \nabla f_i(x_i^{h}) \right \|^2_2 - \frac{\eta}{2}  \left \| \nabla f(\overline{x}^{h}) \right \|^2_2
\end{align}
where the last step comes from the inequality:
\begin{align}
\notag 2 \left \langle \mathbf{a}, \mathbf{b} \right \rangle \le \| \mathbf{a} \|^2_2 + \| \mathbf{b} \|^2_2
\end{align}
for any vectors $\mathbf{a}$, $\mathbf{b} \in \mathbb{R}^d$.

For the third term, we add and subtract $\nabla f(\overline{x}^{h})$ and the sum of $\nabla f_i (x_i^{h})$:
\begin{align} \label{eq:proof-1-3}
&\mathbb{E} \left\| \frac{1}{N}\sum_{i=1}^{N}  \nabla F_i(x_i^{h})  \right\|_2^2 \notag \\
\le & \mathbb{E} \left\| \frac{1}{N} \sum_{i=1}^{N}  \left( \nabla F_i(x_i^{h}) - \nabla f_i(x_i^{h}) \right) \right\|_2^2 \notag \\
+ & \left \| \frac{1}{N}\sum_{i=1}^{N}  \left(\nabla f_i(x_i^{h}) - \nabla f_i(\overline{x}^{h}) + \nabla f_i(\overline{x}^{h})\right) \right \|_2^2 \notag \\
\le & \frac{2 }{N} \sum_{i=1}^{N} \left \|  \nabla f_i(x_i^{h}) - \nabla f_i(\overline{x}^{h}) \right \|_2^2 + 2 \left \| \nabla f(\overline{x}^{h}) \right \|_2^2 + \frac{\sigma^2 }{N}
\end{align}
where the first step comes from the Assumption 2 and the following inequality with $\alpha = 1$:
\begin{align}
\notag \| \mathbf{a} + \mathbf{b} \|^2_2 \le (1+\alpha) \| \mathbf{a} \|_2^2 + (1+\alpha^{-1}) \| \mathbf{b} \|_2^2 , \alpha > 0
\end{align}
for any vectors $\mathbf{a}$, $\mathbf{b} \in \mathbb{R}^d$, and the last step comes from the Assumption 3.

%\begin{align} \label{eq:proof-1-3}
%&\mathbb{E} \left \| \frac{1}{N}\sum_{i=1}^{N} \nabla F_i(x_i^{h}) \right \|_2^2 \notag \\
%= & \left \| \frac{1}{N}\sum_{i=1}^{N} \nabla f_i(x_i^{h}) \right \|_2^2 + \mathbb{E} \left \| \frac{1}{N}\sum_{i=1}^{N} \left (\nabla F_i(x_i^{h})-\nabla f_i(x_i^{h}) \right) \right \|_2^2 \notag \\
%\le & \left \| \frac{1}{N}\sum_{i=1}^{N} \nabla f_i(x_i^{h}) - \nabla f_i(\overline{x}^{h}) + \nabla f_i(\overline{x}^{h}) \right \|_2^2 + \frac{\hat{\sigma}^2}{N} \notag \\
%+ & 
%\end{align}

Combining Eq. \eqref{eq:proof-1-1}, Eq. \eqref{eq:proof-1-2} and Eq. \eqref{eq:proof-1-3} as well as Assumption 1, we obtain:
% \begin{align}
% & \frac{\eta \tau}{2 N} \sum_{i=1}^{N} \left \| \nabla f_i(\overline{x}^{h}) - \nabla f_i(x_i^{h}) \right \|^2_2 - \frac{\eta \tau}{2}  \left \| \nabla f(\overline{x}^{h}) \right \|^2_2 \notag \\
% + & \frac{L \eta^2 \tau^2}{N} \sum_{i=1}^{N} \left \|  \nabla f_i(x_i^{h}) - \nabla f_i(\overline{x}^{h}) \right \|_2^2 \notag \\
% + & L \eta^2 \tau^2  \left \| \nabla f(\overline{x}^{h}) \right \|_2^2 + \frac{L \eta^2 \sigma^2 \tau^2}{2 N}
% \end{align}

\begin{align} \label{eq:proof-1-4}
\mathbb{E}f(\overline{x}^{(h+1)}) & \le f(\overline{x}^{h}) + \eta \tau (L \eta \tau - \frac{1}{2}) \left \| \nabla f(\overline{x}^{h}) \right \|_2^2 \notag \\
& + \frac{\eta L^2 \tau}{N} (\frac{1}{2} + \eta L \tau) \sum_{i=1}^{N} \left \| \overline{x}^{h} - x_i^{h} \right \|_2^2 + \frac{\sigma^2 \eta^2  \tau^2 L}{N}
\end{align}

Applying $\eta \le \frac{1}{4L\tau}$ in the second and the third terms, we complete the proof:
\begin{align} \label{eq:proof-1-5}
\mathbb{E}f(\overline{x}^{(h+1)}) & \le f(\overline{x}^{h})  - \frac{\eta \tau}{4} \left \| \nabla f(\overline{x}^{h}) \right \|_2^2 \notag \\
& + \frac{\eta L^2 \tau}{N}  \sum_{i=1}^{N} \left \| \overline{x}^{h} - x_i^{h} \right \|_2^2 + \frac{\sigma^2 \eta^2  \tau^2 L}{N} %\frac{\sigma^2 }{16 L N}
\end{align}

% \begin{align}
% & \frac{1}{H} \sum_{h=1}^{H} \left \| \nabla f(\overline{x}^{h}) \right \|_2^2 \le \frac{4*(f(\overline{x}^{1})-f(\overline{x}^{*}))}{\eta \tau H} \notag \\
% & + \frac{4 L^2}{N H} \sum_{h=1}^{H}  \sum_{i=1}^{N} \left \| \overline{x}^{h} - x_i^{h} \right \|_2^2 + \frac{4 L \eta \tau \sigma^2}{N}
% \end{align}
\end{IEEEproof}

\begin{remark}
Summing up for all $H$ communication rounds and rearranging the terms in Eq. \eqref{eq:lemma1}, we get:
\begin{align}\label{eq:remark1}
& \frac{1}{H} \sum_{h=1}^{H} \left \| \nabla f(\overline{x}^{h}) \right \|_2^2 \le \frac{4*(f(\overline{x}^{1})-f(\overline{x}^{*}))}{\eta \tau H} \notag \\
& + \frac{4 L^2}{N H} \sum_{h=1}^{H}  \sum_{i=1}^{N} \left \| \overline{x}^{h} - x_i^{h} \right \|_2^2 + \frac{4 L \eta \tau \sigma^2}{N}.
\end{align}
\end{remark}

\begin{lemma}
Under the above assumptions with $\frac{27 L \eta^2}{(1-\rho)^2} < 1$, we have the following formulation:
\begin{align} \label{eq:lemma2}
&\sum_{h=1}^{H} \sum_{i=1}^N \mathbb{E} \left \| \overline{x}^{h} - x^{h}_i\right \|^2_2 
\le \frac{2N\eta^2(\sigma^2 +3\zeta^2)H}{(1-\rho)^2-3\eta^2L^2} \notag \\
& \quad \quad \quad + \frac{6 N \eta^2}{(1-\rho)^2-3\eta^2L^2} \sum_{h=1}^{H} \mathbb{E} \left \| \nabla f(\overline{x}^{h}) \right \|^2_2.
\end{align}
\end{lemma}

\begin{IEEEproof}
Based on the updating rule, we have:
\begin{align} \label{eq:proof-2-1}
X^{h} &= \sum_{s=1}^{h-1} X^{s} W^{h-s} + \sum_{s=1}^{h-1} \eta \nabla F(X^{s}) W^{h-s-1}, \notag \\
\overline{X}^{h} &= \sum_{s=1}^{h-1} X^{s} W^{h-s} \frac{\mathbf{1}}{N} + \sum_{s=1}^{h-1} \eta \nabla F(X^{s}) W^{h-s-1} \frac{\mathbf{1}}{N} \notag \\
& =  \sum_{s=1}^{h-1} \overline{x}^{s} + \sum_{s=1}^{h-1} \eta \overline{\nabla F}(X^{s}).
\end{align}

Thus, we can obtain the following result:
\begin{align} \label{eq:proof-2-2}
& \sum_{i=1}^N \mathbb{E} \left \| \overline{x}^{h} - x^{h}_i\right \|^2_2 \notag \\ 
= & \sum_{i=1}^N \mathbb{E} \left \| \sum_{s=1}^{h-1} \left( X^{s} W^{h-s}\mathbf{e}^{i} - \overline{x}^{s} \right) \right. \notag \\ 
& \left. -  \sum_{s=1}^{h-1} \eta \left( \nabla F(X^{s})W^{h-s-1} \mathbf{e}^{i} - \overline{\nabla F}(X^{s}) \right) \right \|^2_F \notag \\ 
%\le & 2 \sum_{i=1}^{N} \mathbb{E} \left \| \sum_{s=1}^{h-1} \left( Q^{s} W^{h-s}\mathbf{e}^{(i)} - \overline{Q}^{s} \right) \right \|^2  \notag \\
%&+   2 \sum_{i=1}^{N} \mathbb{E} \left \| \sum_{s=1}^{h-1} \eta \left( \nabla F(X^{s})W^{h-s-1} \mathbf{e}^{(i)} - \overline{\nabla F}(X^{s}) \right) \right \|^2 \notag \\
%= & 2 \sum_{i=1}^{N} \sum_{s=1}^{h-1} \mathbb{E} \left \|   Q^{s} W^{h-s}\mathbf{e}^{(i)} - \overline{Q}^{s}  \right \|^2  \notag \\
%& + 2 \sum_{i=1}^{N} \mathbb{E} \left \| \sum_{s=1}^{h-1} \eta \left( \nabla F(X^{s})W^{h-s-1} \mathbf{e}^{(i)} - \overline{\nabla F}(X^{s}) \right) \right \|^2 \notag \\
%& + 4 \sum_{i=1}^{N} \sum_{s \ne s^{\prime}} \mathbb{E} \left \langle Q^{s} W^{h-s} \mathbf{e}^{(i)} - \overline{Q}^{s},  Q^{(s^{\prime})} W^{h-s^{\prime}} \mathbf{e}^{(i)} - \overline{Q}^{(s^{\prime})} \right \rangle \notag \\ 
\le & 2 \sum_{i=1}^{N} \sum_{s=1}^{h-1} \mathbb{E} \left \|   X^{s} W^{h-s}\mathbf{e}^{i} - \overline{x}^{s}  \right \|^2_F  \notag \\
& + 2 \sum_{i=1}^{N} \mathbb{E} \left \| \sum_{s=1}^{h-1} \eta \left( \nabla F(X^{s})W^{h-s-1} \mathbf{e}^{i} - \overline{\nabla F}(X^{s}) \right) \right \|^2_F \notag \\
\le & 2 \mathbb{E} \sum_{s=1}^{h-1} \left \| \rho^{h-s} X^{s} \right \|_F^2 + 2 \mathbb{E} \left( \sum_{s=1}^{h-1} \eta \rho^{h-s-1} \left \| \nabla F(X^{s})\right \|_F \right)^2
\end{align}

To bound the term $\left \| \nabla F(X^{s})\right \|_F^2$ in Eq. \eqref{eq:proof-2-2}, we first bound $\left \| \nabla F_i(x_i^{h}) \right \|_2^2$ as follows:
\begin{align} \label{eq:proof-2-4}
& \mathbb{E} \left \| \nabla F_i(x_i^{h}) \right \|_2^2 \nonumber \\
= & \mathbb{E} \left \| \nabla F_i(x_i^{h}) - \nabla f_i(x_i^{h}) + \nabla f_i(x_i^{h}) \right \|_2^2 \nonumber \\
= & \mathbb{E} \left \| \nabla F_i(x_i^{h}) - \nabla f_i(x_i^{h})\right \|_2^2 + \mathbb{E} \left \| \nabla f_i(x_i^{h}) \right \|_2^2 \nonumber \\
& + 2 \mathbb{E} \left \langle \mathbb{E} \nabla F_i(x_i^{h}) - \nabla f_i(x_i^{h}), \nabla f_i(x_i^{h}) \right \rangle \nonumber \\
= & \mathbb{E} \left \| \nabla F_i(x_i^{h}) - \nabla f_i(x_i^{h})\right \|_2^2 + \mathbb{E}\left \| \nabla f_i(x_i^{h}) \right \|_2^2 \nonumber \\
\le & \sigma^2 + \mathbb{E} \left \| \nabla (f_i(x_i^{h}) - \nabla f_i(\overline{x}^{h})) \right. \nonumber \\
& \left. + (\nabla f_i(\overline{x}^{h}) - \nabla f(\overline{x}^{h})) + \nabla f(\overline{x}^{h})\right \|_2^2 \nonumber \\
\le & \sigma^2 + 3 \mathbb{E} \left \| \nabla f_i(x_i^{h}) - \nabla f_i(\overline{x}^{h})\right \|_2^2  \nonumber \\
& + 3 \mathbb{E} \left \| \nabla f_i(\overline{x}^{h}) - \nabla f(\overline{x}^{h}) \right \|_2^2 + 3 \mathbb{E} \left \| \nabla f(\overline{x}^{h})\right \|_2^2 \nonumber \\
\le & \sigma^2 + 3L^2 \mathbb{E} \left \|\overline{x}^{h} - x_i^{h} \right \|_2^2 + 3 \zeta^2 + 3 \mathbb{E} \left \| \nabla f(\overline{x}^{h})\right \|_2^2 
\end{align}
which means
\begin{align}
\mathbb{E} \left \| \nabla F(X^{h})\right \|_F^2 &\le \sum_{i=1}^{N} \left \| \nabla F_i(x_i^{h}) \right \|_2^2 \nonumber \\
&\le  N \sigma^2 + 3 L^2 \sum_{i=1}^N \mathbb{E} \left \| \overline{x}^{h} - x^{h}_i\right \|^2_2 \nonumber \\ & \quad \quad + 3 N \zeta^2  + 3 N \mathbb{E} \left \| \nabla f(\overline{x}^{h})\right \|_2^2
\end{align}

Inserting Eq. \eqref{eq:proof-2-4} into Eq. \eqref{eq:proof-2-2}, applying Lemmas 5 and 6  in \cite{tang2018communication}, and setting $\frac{3L^2\eta^2}{(1-\rho)^2} < 1$, we complete the proof:
\begin{align} \label{eq:proof-2-3}
&\sum_{h=1}^{H} \sum_{i=1}^N \mathbb{E} \left \| \overline{x}^{h} - x^{h}_i\right \|^2_2 
\le \frac{2N\eta^2(\sigma^2 +3\zeta^2)H}{(1-\rho)^2-3\eta^2L^2} \notag \\
& \quad \quad \quad + \frac{6 N \eta^2}{(1-\rho)^2-3\eta^2L^2} \sum_{h=1}^{H} \mathbb{E} \left \| \nabla f(\overline{x}^{h}) \right \|^2_2
\end{align}

\end{IEEEproof}

\begin{remark}
Inserting Eq. \eqref{eq:lemma2} into Eq. \eqref{eq:remark1}, we obtain the following convergence bound:
\begin{align} \label{eq:convergence bound}
& \frac{1}{H} \sum_{h=1}^{H} \left \| \nabla f(\overline{x}^{h}) \right \|_2^2 \le \frac{4(f(\overline{x}^{1})\!-\!f(\overline{x}^{*}))((1\!-\!\rho)^2\!-\!3\eta^2L^2)}{\eta \tau H ((1\!-\!\rho)^2\!-\!27\eta^2L^2)} \notag \\
& + \frac{8 L^2 \eta^2(\sigma^2 +3\zeta^2)}{(1-\rho)^2-27\eta^2L^2} + \frac{(1-\rho)^2-3\eta^2L^2}{(1-\rho)^2-27\eta^2L^2} \frac{4 L \eta \tau \sigma^2}{N}.
\end{align}
\end{remark}
The communication topology weight matrix $W$ (reflected by $\rho$), local updating frequency $\tau$ and data distribution (reflected by $\zeta$) all have impacts on the convergence rate with Eq. \eqref{eq:convergence bound}. 
The sparser the topology is, the larger $\rho$  is.
For example, $\rho$ is 0 for the fully-connected topology while  $\rho$ is 0.99 for the ring topology with 36 workers.
Thus, with the increasing of topology sparsity, the above convergence bound will increase.
When $\tau \le \sqrt{\frac{N (f(\overline{x}^{1})-f(\overline{x}^{*}))}{L H \eta^2 \sigma^2}}$, the above convergence bound will decrease as local updating frequency $\tau$ increases.
On the contrary, when $\tau>\sqrt{\frac{N (f(\overline{x}^{1})-f(\overline{x}^{*}))}{L H \eta^2 \sigma^2}}$, the trend of convergence bound and local updating frequency is opposite.
% the above convergence bound will \bluenote{increase with the increasing of local updating frequency $\tau$}.
As the degree of non-IID data distribution increases (\ie, larger $\zeta$), the upper bound of Remark 2 will get looser and looser.

\begin{figure}[t]
	\centering
	%		\vspace{5pt}
	\subfigure[Accuracy \vs $\tau$]
	{
		\includegraphics[width=0.45\linewidth,height=3.1cm]{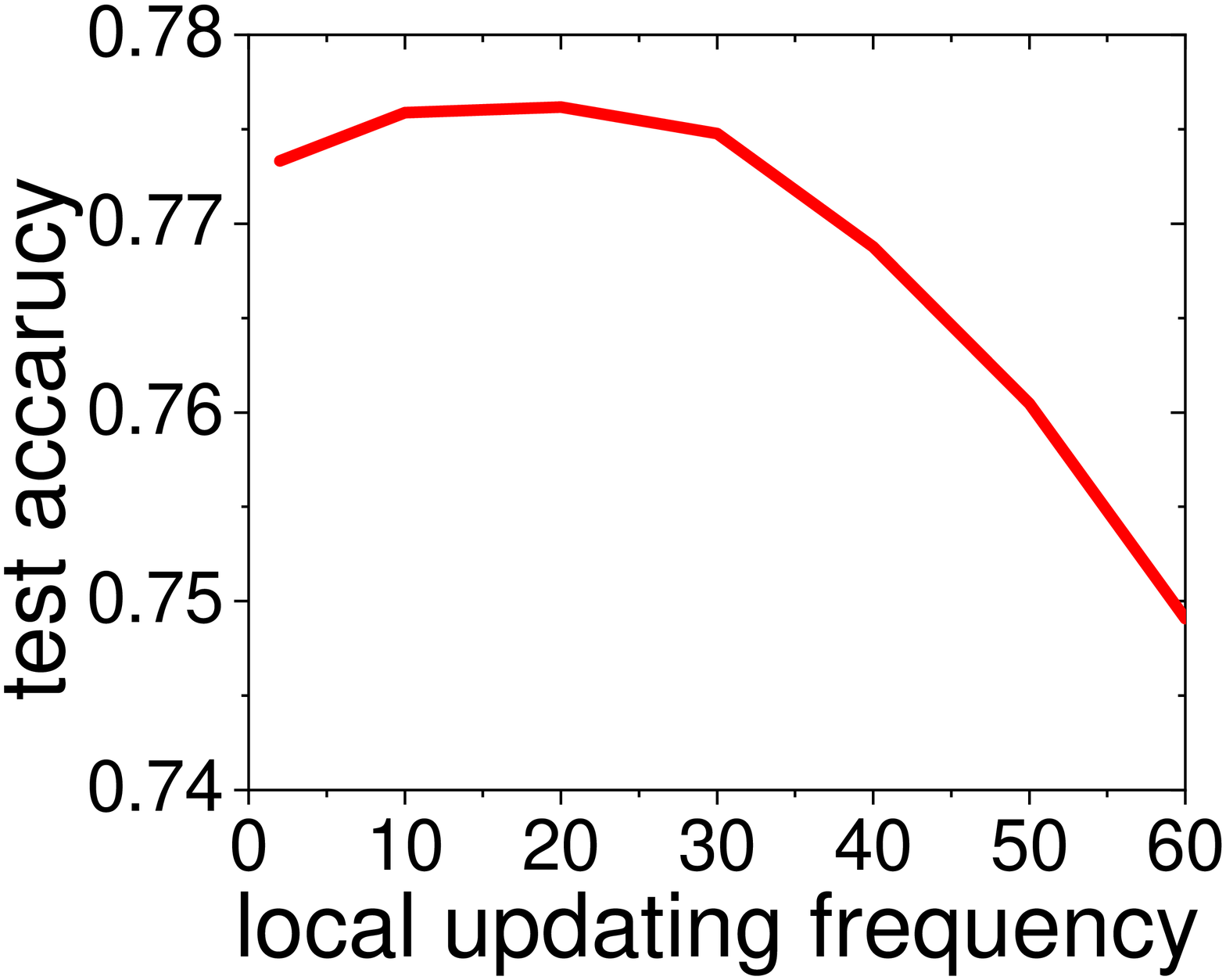}
        \label{fig:tau_acc}
	}
	\subfigure[Completion time \vs $\tau$]
	{
		\includegraphics[width=0.45\linewidth,height=3.1cm]{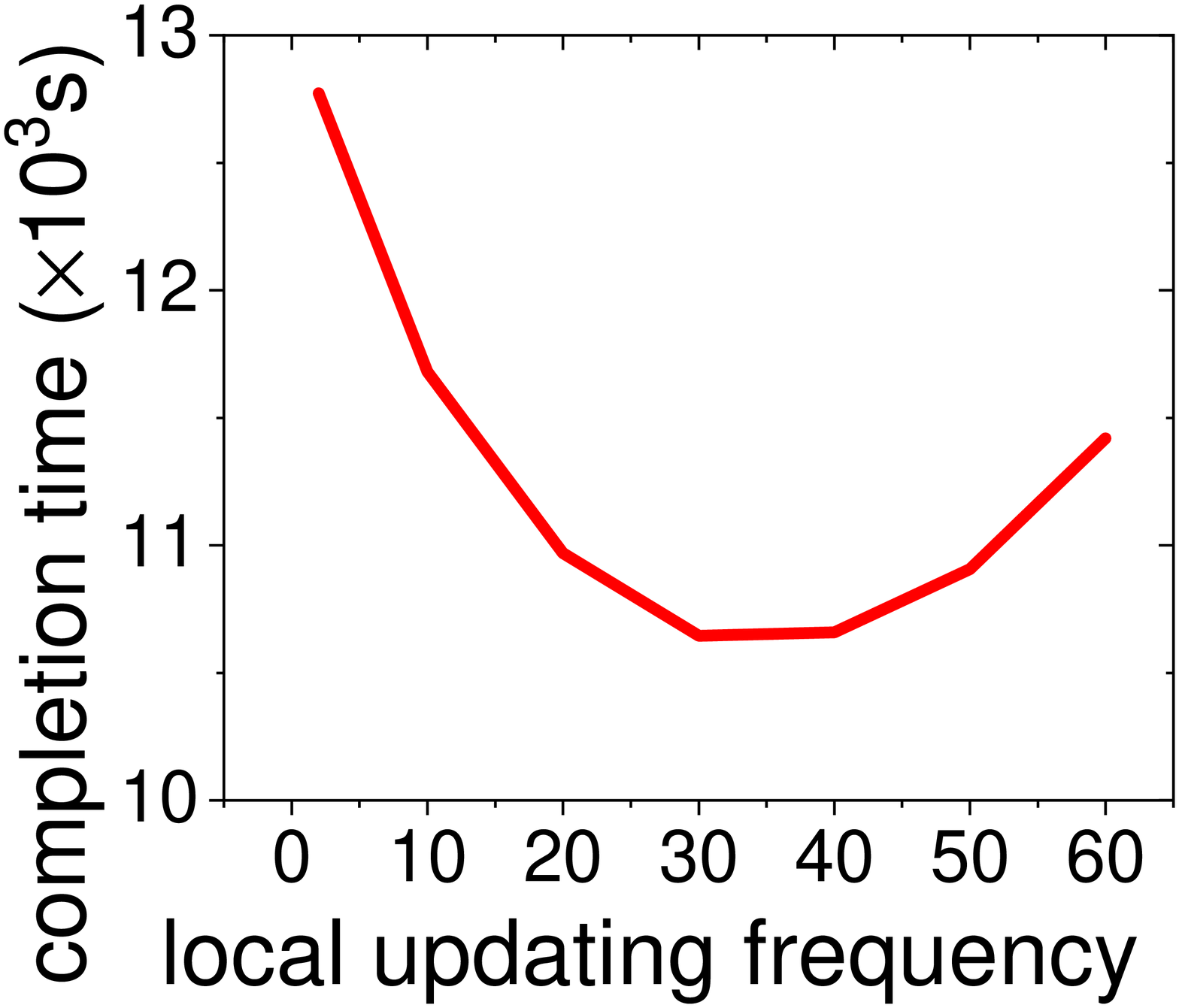}
		\label{fig:tau_time}
	}
	\caption{Model training on different local updating frequency $\tau$ of CIFAR-10.}
	\label{fig:tau}
	\vspace{-1.2em}
\end{figure}

According to the above analysis, a very large local updating frequency may make the decentralized models converge to the local optimal solutions rather than the global optimum.
However, a relatively smaller local updating frequency will lead to more communication rounds until convergence, incurring more computing time and communication time.
To observe the impact of local updating frequency on model training, we conduct a pre-experiment for training AlexNet on CIFAR-10 and record the model accuracy and completion time with different local updating frequencies.
As shown in Fig. \ref{fig:tau_acc}, the model accuracy decreases with increasing of local updating frequency when $\tau > 27$.
% ($c_1$ is a constant).
% That is because a very large local updating frequency may make the decentralized models converge to the local optimal solutions rather than the global optima.
% However, a relatively smaller local updating frequency will lead to more communication rounds until convergence, incurring more computing time and communication time.
Besides, Fig. \ref{fig:tau_time} shows that the completion time of model training decreases with increasing of local updating frequency when $\tau < 30$.
These results are consistent with our analysis in Eq. \eqref{eq:convergence bound}.
% ($c_2$ is a constant).
Therefore, it is critical to determine the appropriate local updating frequencies for different workers to accelerate model training.

% \begin{theorem}
% Based on Lemma 1, we have the following convergence bound:
% \begin{align}
% &\frac{1}{H} \sum_{h=1}^{H} \mathbb{E} \|\nabla f(\overline{x}^{h}) \|^2_2 \le \frac{\sigma(1+\tau)}{\sqrt{N \tau H}} \notag\\
% &+ \frac{1}{(1-\rho)^2} (\frac{\zeta}{H})^{\frac{2}{3}} + \frac{N}{H \tau(1-\rho)^2}
% \end{align}
% where the local learning rate $\eta$ satisfy
% \begin{equation}
%     \eta = \left(\frac{6L}{\sqrt{(1-\rho)^2}} + \sigma N^{-\frac{1}{2}}\tau^{\frac{1}{2}}H^{\frac{1}{2}} + \zeta^{\frac{2}{3}}H^{\frac{1}{3}}\right)^{-1}
% \end{equation}
% \end{theorem}

\begin{corollary}\label{cly}
Let the local learning rate $\eta$ satisfy the following constraint: 
\begin{equation}
    \eta = (\frac{6L}{\sqrt{(1-\rho)^2}} + \sigma N^{-\frac{1}{2}}\tau H^{\frac{1}{2}} + \zeta^{\frac{2}{3}}H^{\frac{1}{3}})^{-1}.
\end{equation}
The convergence upper bound can be transformed as:
\begin{align}
&\frac{1}{H} \sum_{h=1}^{H} \mathbb{E} \|\nabla f(\overline{x}^{h}) \|^2_2 \le \frac{\sigma}{\sqrt{N H}} \notag\\
&+ \frac{1}{(1-\rho)^2} (\frac{\zeta}{H})^{\frac{2}{3}} + \frac{1}{H \tau^2 (1-\rho)^2}.
\end{align}
% where the local learning rate $\eta$ satisfy
% \begin{equation}
%     \eta = \left(\frac{6L}{\sqrt{(1-\rho)^2}} + \sigma N^{-\frac{1}{2}}\tau H^{\frac{1}{2}} + \zeta^{\frac{2}{3}}H^{\frac{1}{3}}\right)^{-1}
% \end{equation}
\end{corollary}

With Corollary \ref{cly}, our method can achieve a linear speedup of convergence rate $\mathcal{O}(\frac{1}{\sqrt{H N}})$ as stated in many previous works \cite{koloskova2019decentralized, lian2017can}, indicating that our method will contribute to speeding up the training without loss of convergence performance.

\section{Algorithm Design}\label{sec:alg}
% In this section, we first show how to estimate $D_i^{h}$  using consensus distances of the connected workers.
% Then, we transform the non-linear mixed integer programming problem in Eq. \eqref{problem} into a linear programming, and propose an efficient algorithm to solve the joint optimization of local updating frequency and network topology construction.

\subsection{Consensus Distance Estimation}\label{subsec_distance_estimation}
We first analyze how the network topology and local updating frequency affect the consensus distance between the model of worker $i$ and the average of all workers' models.
According to the update rule in Eq. (\ref{Eq:Update rule 1}) and the definition in Eq. (\ref{Eq:Local Consenus Distance}), the consensus distance $\| \overline{x}^{h+1} - x_i^{h+1}\|_2$ at round $h+1$ can be formulate as:
% between worker $i$ and the average of all workers' models
% \begin{scriptsize}
\begin{align}
&D^{h+1}_{i} = \| \overline{x}^{h+1} - x_i^{h+1}\|_2 \notag\\
& = \left\| \frac{1}{N}\sum_{j=1}^N x_j^{h, \tau_j^h} - ( x_i^{h, \tau_i^h} + w_{i,j}^{h} \sum_{j=1}^N a_{i,j}^{h} (x_j^{h, \tau_j^h} - x_i^{h, \tau_i^h} ) ) \right\|_2 \notag\\
& = \left\| \sum_{j=1}^N (\frac{x_j^{h, \tau_j^h} - x_i^{h, \tau_i^h}}{N} - w_{i,j}^{h} a_{i,j}^{h} (x_j^{h, \tau_j^h} - x_i^{h, \tau_i^h} ) ) \right\|_2.
\end{align}
% \end{scriptsize}
According to $w_{i,j}^{h} = \frac{1}{u_{max}^{h} + 1}$ in Eq. \eqref{Eq:Step size}, we set $u_{max}^{h} = N-1$ for simplicity, which is the possible maximum value \cite{xiao2004fast}.
Thus, it follows:
\vspace{-0.5em}
\begin{align}\label{Eq:node-pair-dist}
\mathbb{E} D^{h+1}_{i} &= \left\| \sum_{j=1}^N \frac{(1 - a_{i, j}^{h})(x_j^{h, \tau_j^h} - x_i^{h, \tau_i^h})}{N} \right\|_2 \notag\\
& \le \frac{1}{N}\sum_{j=1}^N (1 - a_{i, j}^{h})  D_{i, j}^{h}\mbox{,} 
\end{align}
where $D_{i, j}^{h} = \| x_i^{h, \tau_i^h} - x_j^{h, \tau_j^h}\|_2$ ($\forall i, j \in [N] $) is the consensus distance between two models of worker $i$ and worker $j$. The last step of Eq. \eqref{Eq:node-pair-dist} follows the triangle inequality.
After receiving local models of neighbors, worker $i$ can locally calculate the consensus distance $D_{i, j}^{h}$, $\forall j \in \mathcal{N}_i^{h}$.
%  between worker $i$ and each neighbor $j$
% Worker $i$ calculates $D_{i, j}^{h}, \forall j \in \mathcal{N}_i^{h}$, locally after receiving neighbor $j$'s model.
As a result, the upper bound of the average consensus distance in Eq. \eqref{Eq:Avg Consenus Distance} can be expressed as:
\vspace{-0.6em}
\begin{equation} \label{Eq:Distance upper bound}
% \begin{split}
\mathbb{E} D^{h+1} \le \frac{1}{N^2} \sum_{i=1}^N \sum_{j=1}^N (1 - a_{i, j}^{h}) D_{i, j}^{h}.
% \end{split}
\end{equation}
% \vspace{-0.3em}
Note that when we set $a_{i, j}^{h} = 1$, $\forall i, j \in [N]$, the upper bound of average consensus distance $D^{h+1}$ is 0, \ie, if each worker receives local models from all others, the updated models among workers are identical.
% Note that when we set $a_{i, j}^{k} = 1, \forall i, j \in [N]$, it is easy to obtain the upper bound for the PS-based training, where the average consensus distance $D^{h+1}$ is 0 after model aggregation, \ie, if each worker receives local models from all others, the models among workers are identical.

To solve the problem in Eq. \eqref{problem} with Eq. \eqref{Eq:Distance upper bound}, we still need to know the consensus distances among models of all workers.
%At training round $k$, before exchanging models, we don't know node pairs' consensus distance, so we use the parameters received at last round to estimate current nodes pair's consensus distance.
However, if worker $i$ and worker $j$ are not connected at round $h$, it is infeasible to obtain their consensus distance directly since each worker only receives local models from its neighbors.
Thus, we need to estimate the consensus distance between unconnected workers with the help of those of the connected workers.
% \bluenote{Every worker would send the consensus distance calculated by itself to the coordinator at each communicating round.}
Firstly, when the coordinator has collected consensus distance $D_{i, p}^{h}$ and $D_{p, j}^{h}$, $\forall p \in {N} \setminus \{i, j\}$, $D_{i, j}^{h}$ can be estimated as:
% \vspace{-0.3cm}
\begin{align} \label{Eq:consensus estimation 2}
D_{i, j}^{h} &=  \left \| x_i^{h, \tau_i^h} - x_p^{h, \tau_p^h} + x_p^{h, \tau_p^h} - x_j^{h, \tau_j^h} \right \|_2 \notag\\
&\le \left \| x_i^{h, \tau_i^h} - x_p^{h, \tau_p^h} \right \|_2 +  \left \| x_p^{h, \tau_p^h} - x_j^{h, \tau_j^h} \right \|_2 \notag\\
&= D_{i, p}^{h} + D_{p, j}^{h}\mbox{,}
\end{align}
where the second step follows the triangle inequality.
Thus we can estimate $D_{i, j}^{h}$ as $\hat{D}_{i, j}^{h}$:
\begin{equation} \label{Eq:minimum dist}
\hat{D}_{i, j}^{h} = \min_{p \in [N]  \setminus \{i, j\}} (D_{i, p}^{h} + D_{p, j}^{h}).
\end{equation}

Secondly, if there is no common neighbor between worker $i$ and worker $j$ at round $h$ (\ie, $\mathcal{N}_i^{h} \cap \mathcal{N}_j^{h} = \emptyset$), we can use Eq. (\ref{Eq:consensus estimation 2}) and Eq. (\ref{Eq:minimum dist}) iteratively to obtain $\hat{D}_{i, j}^{h}$.
Since the network topology is a connected graph, the above problem is equivalent to the shortest path problem, which can be solved efficiently by the Floyd-Warshall algorithm \cite{black1998dictionary} at the coordinator.
As the triangle inequality may amplify consensus distance among workers, the historical consensus distance is used to make our estimation more stable and accurate.
Specifically, we use the exponential moving average to smooth the consensus distance, with $\beta_1 \in [0, 1]$, as follows:
\begin{equation} \label{Eq:consensus estimation 3}
D_{i,j}^{h} = (1 - \beta_1)D_{i,j}^{h-1} + \beta_1 \hat{D}_{i,j}^{h},\ \text{if}\ a_{i, j}^{h} = 0.
\end{equation}

\subsection{Algorithm Description} \label{sec:alg1}
Firstly, to minimize the average waiting time of all workers, we let the $t_i^h$ among workers be approximately equal.
Then we can have the following formulation:
% \begin{equation}\label{tau_max}
%     \tau_{i}^{h} = \lfloor \tau_{l}^{h} \cdot \frac{\mu_l^h+\max \{\beta_{l,j}^h\}}{\mu_i^h+\max \{\beta_{i,j}^h\}} \rfloor
% \end{equation}
\begin{equation}\label{tau_max}
    \lfloor \frac{\tau_{l}^{h} \cdot \mu_l^h+\max \{\beta_{l,j}^h\}}{\tau_{i}^{h} \cdot\mu_i^h+\max \{\beta_{i,j}^h\}} \rfloor = 1\mbox{,}
\end{equation}
where $l$ denotes the index of the fastest worker with the largest local updating frequency at round $h$.
Thus, $\tau = \tau_{l}^{h}$.
Then the total training time can be formulated as follows:
% \vspace{-0.1em}
\begin{equation}\label{eq:round_time}
	T(H,\tau)=\sum_{h=1}^H (\tau \cdot \mu_l^h+ \max \{\beta_{l,j}^h\}).
\end{equation}

Secondly, the problem in Eq. \eqref{problem} is a non-linear mixed integer programming problem, which is hard to solve \cite{karp1972reducibility, papadimitriou1982complexity}.
However, given a specific network topology, we can take the upper bound of $D^{h+1}$ in Eq. (\ref{Eq:Distance upper bound}) as the estimation and transform Eq. \eqref{problem} into a linear programming problem as:
% \bluenote{However, if we determined the network topology and take the upper bound of $D^{(k+1)}$ in Eq. (\ref{Eq:Distance upper bound}) as the estimation, then Eq. \eqref{problem} can be transformed into a linear programming as follows:}

\vspace{0.1cm}
\centerline{$\min T(H,\tau)$}
\vspace{-0.3cm}
\begin{equation}\label{eq:DFLproblem-refor}
s.t.
\begin{cases}
	\frac{1}{N^2} \sum_{i=1}^N \sum_{j=1}^N (1 - a_{i, j}^{h}) D_{i, j}^{h} \le D_{max}^h \vspace{2mm}\\
	\lfloor \frac{\tau_{l}^{h} \cdot \mu_l^h+\max \{\beta_{l,j}^h\}}{\tau_{i}^{h} \cdot\mu_i^h+\max \{\beta_{i,j}^h\}} \rfloor = 1
\end{cases}
\end{equation}

In terms of Eq. \eqref{eq:DFLproblem-refor}, we propose an efficient algorithm, that adaptively determines local updating frequency for each worker and constructs the network topology.
And the coordinator is responsible for monitoring the network condition and recording the model training status.
% To monitor the dynamics of network and coordinate workers' training and communication, as suggested in \cite{kairouz2019advances}, a coordinator is still required. 

We present the procedure for workers (Alg. \ref{alg:clients}) and the coordinator (Alg. \ref{alg:coordinator}) while the proposed algorithm is formally described in Alg. \ref{alg:algorithm}.
In Alg. 1, at the beginning of round $h$, each worker $i$ requests the information about its neighbor set $\mathcal{N}_i^{h}$ and local updating frequency $\tau_i^h$ from the coordinator.
Then worker $i$ performs local updating of $\tau_i^h$ times by Eq. \eqref{Eq:Update rule 2} and estimates the parameters $L_{i}$ and $\sigma_i$.
After local updating is finished, worker $i$ sends the local model to its neighbors and waits for receiving the models from its neighbors for aggregation.
The local updating frequency of each worker is associated with its computing and communicating capabilities.
For instance, the workers with high performance are assigned with larger local updating frequencies, so that each worker does not need to waste too much waiting time.
% \bluenote{The local updating frequency of each worker is associated with its capabilities of computing and communication, so that each worker does not need to waste too much waiting time.}
% Worker $i$ would also record computing time $\mu_i^{h}$ and communication time $\beta_{i,j}^{h}$ between worker $i$ and worker $j$.
% By comparing local model with receiving model(s), 
After receiving models from the neighbors, worker $i$ computes consensus distance $D_{i, j}^{h}$, $\forall j \in \mathcal{N}_i^{h}$.
Finally, worker $i$ sends network conditions, model training statuses, and other parameters to the coordinator and starts the next communication round.
% parameters $L_i$, $\sigma_i$, consensus distance $D_{i, j}^{h}$ and the recording time $\mu_i^{h}$, $\beta_{i,j}^{h}$ to the coordinator and starts the next communication round.

In Alg. \ref{alg:coordinator}, the coordinator waits for receiving the parameters (\ie, $L_i$ and $\sigma_i$), consensus distance (\ie, $D_{i, j}^{h}$), computing time (\ie, $\mu_i^{h}$) and communication time (\ie, $\beta_{i,j}^{h}$) from workers, and takes average of parameters $L_i$ and $\sigma_i$ to get $L$ and $\sigma$.
Then the coordinator calls Alg. \ref{alg:algorithm} to get local updating frequencies and network topology of different workers for the next communication round.
% At the coordinator side, Alg. \ref{alg:coordinator} starts with sending the initial neighbors and local updating frequency of each worker to workers.
% Then the coordinator waits for receiving the parameters $L_i$, $\sigma_i$, consensus distance $D_{i, j}^{h}$ and the recording time $\mu_i^{h}$, $\beta_{i,j}^{h}$ from each worker and takes average of the parameters $L_i$, $\sigma_i$.
% After that, the coordinator calls Alg. \ref{alg:algorithm} to get the network topology and local updating frequency of each worker for next communication round.

\begin{algorithm}[!t]
	\caption{Procedure at worker $i$}\label{alg:clients}
	% \algorithmicensure{ $x_i^{H}$}
	\begin{algorithmic}[1]
		\For {$h=1$ to $H$}
		\State Receive $\mathcal{N}_i^{h}$ and $\tau_i^h$ from the coordinator;
		\State Perform local updating of $\tau_i^h$ times by Eq. \eqref{Eq:Update rule 2};
		\State {Estimate $L_{i} \leftarrow \frac{\|\nabla f_{i}(x_i^{h+1})-\nabla f_{i}(x_i^{h})\|}{\|x_i^{h+1}-x_i^{h}\|}$};
		\State {Estimate $\sigma_{i} \leftarrow \mathbb{E}\left[\|\nabla F_{i}(x_i^h, \xi_{i}^{h})-\nabla f_{i}(x_i^h)\|^{2}\right]$};
		\State Send local model to workers in $\mathcal{N}_i^{h}$;
		\State Receive models from workers in $\mathcal{N}_i^{h}$;
		\State Aggregate models by Eq. (\ref{Eq:Update rule 1}) and obtain $x_i^{h+1}$;
		\State Record computing time $\mu_i^{h}$ and communication time $\beta_{i,j}^{h}$, $\forall j \in \mathcal{N}_i^{h}$;
		\State Compute consensus distance $D_{i, j}^{h}$, $\forall j \in \mathcal{N}_i^{h}$;
		% \State {$\sigma_{i}\leftarrow\mathbb{E}\left[\|\nabla F_{i}(\boldsymbol{w}^{\hat{h}}, \xi_{i}^{\hat{h}})-\nabla F_{i}(\boldsymbol{w}^{\hat{h}})\|^{2}\right]$}
		\State Send $\mu_i^{h}$, $\beta_{i,j}^{h}$, $D_{i, j}^{h}$, $L_i$, $\sigma_i$ to the coordinator;
		\EndFor
	\end{algorithmic}
	\begin{flushleft}
		{\bf Output:}
		$x_i^{H}$.
	\end{flushleft}
\end{algorithm}

\begin{algorithm}[!t]
	\caption{Procedure at coordinator}\label{alg:coordinator}
	%\algorithmicensure{ $\boldsymbol{w}^{H}$}
	\begin{algorithmic}[1]
		\For {$h=1$ to $H$}
		\State Send $\mathcal{N}_i^{h}$ and $\tau_i^h$ to worker $i$, $\forall i \in [N]$;
		\State Receive $\mu_i^{h}$, $\beta_{i,j}^{h}$, $D_{i, j}^{h}$, $L_i$, $\sigma_i$ from worker $i$, $\forall i \in [N]$;
		\State $L \leftarrow \frac{1}{N} \sum_i^N L_i$;
		\State $\sigma \leftarrow \frac{1}{N} \sum_i^N \sigma_i$;
		\State Determine the local updating frequency and network topology for each worker by the proposed algorithm in Alg. \ref{alg:algorithm};
		\EndFor
	\end{algorithmic}
\end{algorithm}

\begin{algorithm}[t]
	\caption{Adaptive control algorithm of FedHP}\label{alg:algorithm}
	%\algorithmicensure{ $\boldsymbol{w}^{H}$}
	%\algorithmicrequire{$D_{i,j}^h$}
	\begin{flushleft}
		{\bf Input:} $\mu_i^h$, $D_{i,j}^h$, $\beta_{i,j}^h$, $\forall i, j \in [N]$; $L$, $\sigma$; $D_{max}^{h}$; $\mathbf{A}_{b}$.
	\end{flushleft}
	\begin{algorithmic}[1]
		\State Initialize adjacent matrix $\mathbf{A}^{h} = \mathbf{A}_{b}$, search step $s=N$ and $Flag = True$;
		%\State $L \leftarrow \frac{1}{N} \sum_i^N L_i$, $\sigma \leftarrow \frac{1}{N} \sum_i^N \sigma_i$;
		\State Minimize $T_i(H, \tau_i^h) = \sum_{h=1}^H (\tau_{i}^{h} \cdot \mu_i^h+ \max \{\beta_{i,j}^h\})$ and abtain $T_i$ and $\tau_i^h$ of worker $i$, $\forall i\in [N]$;
		\State $l \leftarrow \arg \min_{i} (T_i)$, $T \leftarrow T_l$ and $\tau \leftarrow \tau_l$;
		% \State Search $\tau_{l}^{h}\in[\tau_{s}, \tau_{e}]$ to minimize $T(H,\tau_{l}^{h})$ and abtain the 
		% \State Calculate $T=\sum_{h=1}^H (\tau_{l}^{h} \cdot \mu_l^h+ \max \{\beta_{l,j}^h\})$;
		\While{$True$}
		\If{$Flag$}
		\State $s = \lfloor \sqrt{\sum_{i,j}a_{i,j}^{h}} \rfloor$;
		\Else
		\State $s = \lfloor s / 2 \rfloor$;
		\EndIf
		\State Select $s$ slowest links under the threshold of Eq. \eqref{eq:DFLproblem-refor} into $E$;
		\State Initialize $\mathbf{A}^{\prime} \leftarrow \mathbf{A}^{h}$;
		\For{each link $e_{i,j} \in E$}
		\State Set $a_{i, j} \in \mathbf{A}^{\prime}$ as $0$;
		\If{$\mathbf{A}^{\prime}$ is not connected}
		\State Set $a_{i, j} \in \mathbf{A}^{\prime}$ as $1$;
		\EndIf
		\EndFor
		\State Minimize $T_i(H, \tau_i^h) = \sum_{h=1}^H (\tau_{i}^{h} \cdot \mu_i^h+ \max \{\beta_{i,j}^h\})$ and abtain $T_i$ and $\tau_i^h$ of worker $i$, $\forall i\in [N]$;
		\State $l^{\prime} \leftarrow \arg \min_{i} (T_i)$, $T^{\prime} \leftarrow T_{l^{\prime}}$ and  $\tau \leftarrow \tau_{l^{\prime}}$;
		\If{$T^{\prime} < T$}
		% \State $l \leftarrow l^{\prime}$, $T \leftarrow T^{\prime}$, $\tau \leftarrow \tau_{l^{\prime}}$, $\mathbf{A}^{h} \leftarrow \mathbf{A}^{\prime}$, $Flag \leftarrow True$;
		\State $l$, $T$, $\tau$, $\mathbf{A}^{h}$, $Flag$ $\leftarrow$ $l^{\prime}$,$T^{\prime}$, $\tau_{l^{\prime}}$, $\mathbf{A}^{\prime}$ ,$True$;
		\Else
		\State $Flag \leftarrow False$;
		\EndIf
		\If{not $Flag$ and $s==1$}
		\State Break;
		\EndIf
		\EndWhile
		\State Calculate $\tau_i^h$ for each worker by Eq. \eqref{tau_max}, where $\tau_l^h=\tau$;
	\end{algorithmic}
	\begin{flushleft}
		{\bf Output:}
		$\tau_i^h$, $\forall i \in [N]$, $\mathbf{A}^{h}$.
	\end{flushleft}
\end{algorithm}

As indicated in Eq. \eqref{eq:round_time}, the completion time of model training depends on the slowest link and the slowest worker.
Thus we mainly use the greedy algorithm to remove the slow links in the current network topology to reduce the completion time under the threshold of consensus distance in Eq. \eqref{eq:DFLproblem-refor}.
The procedure executes iteratively until the completion time cannot be reduced after removing any slow links.
Specifically, we take the network conditions, model training statuses of workers, and other parameters as the algorithm input.
Firstly, we start from the base topology (\ie, $\mathbf{A}_{b}$) which includes all available links for P2P communication.
Then we set $\tau_i^h=\sqrt{\frac{N f(\overline{x}^{1})}{L H \eta^2 \sigma^2}}$ and minimize $T_i(H,\tau_i^h)$ by using an LP solver to obtain $T_i$ and $\tau_i^h$ for worker $i$, $\forall i \in [N]$.
We obtain the minimum of completion time $T_l$ in the base topology and get the local updating frequency $\tau_l^h$ of worker $l$ at round $h$ (Line 1-3), where $l=\arg \min_{i} (T_i)$. 
In order to search the optimal topology and local updating frequencies efficiently, we first take a large search step.
Concretely, we set the search step $s$ as the square root of the number of links in the current topology (Line 5-6).
At round $h$, since the slow links may become the system bottleneck in terms of time, we use a greedy algorithm to remove $s$ slowest links and obtain the new network topology $A^{\prime}$ (Line 10-14).
Then we minimize $T_i(H,\tau_i^h)$ again to obtain the new minimum of completion time $T_{l^{\prime}}$ in the new topology and get the new local updating frequency $\tau_{l^{\prime}}$ (Line 15-16).
% Then we minimize $T_i(H,\tau_i)$ again to obtain new $T_i$ and $\tau_i^h$ of worker $i$, $\forall i \in [N]$.
% After that, we obtain the new minimum of completion time $T_{l^{\prime}}$ of the new topology and get the new local updating frequency $\tau_{l^{\prime}}$ (Line 15-16).
If a better solution (\ie, shorter completion time) is found, the current network topology and local updating frequency are updated (Line 17-18).
If we cannot find a better solution at the current search step, the search step is reduced by half.
If the completion time $T$ cannot be further reduced by removing any link, we stop searching and obtain the final network topology as well as local updating frequency of worker $l$.
It is worth noting that we only remove the links that will not affect the connectivity of the network topology and exceed the constraint of consensus distance $D_{max}^{h}$ in Eq. \eqref{eq:DFLproblem-refor}.
% To satisfy the constraint of Eq. \eqref{problem}, we only remove the links that will not affect the connectivity of the network topology (Line 15-18).
% As analyzed in Remark 1, if the average consensus distance does not exceed a critical threshold, the models in DFL can converge as fast as the centralized training.
% However, in the decentralized training, it is infeasible to obtain the upper bound in Remark 1.
In our algorithm, we follow \cite{lin2021on} to set the threshold of $D_{max}^{h}$ adaptively.
%Generally, at the beginning of training, we provide a large $D_{max}^{k}$ and gradually decrease it to make sure workers reach consensus. As suggested in \cite{lin2021on}
Specifically, $D_{max}^{h}$ is the exponential moving average of the gradient norm:
\begin{equation}\label{eq:ema-norm}
D_{max}^{h} = (1-\beta_2) D_{max}^{h-1} + \frac{\beta_2}{N }\sum_{i=1}^N \left\| g_i^{h}\right\|_2\mbox{,}
\end{equation}
where $\frac{1}{N}\sum_{i=1}^{N} \left\| g_i^{h}\right\|_2$ denotes the average norm of local updates at round $h$ among all workers and $\beta_2 \in [0, 1]$.

Herein, we analyze the time complexity of Alg. \ref{alg:algorithm}. As described above, the proposed algorithm reduces the search step $s$ by half if a better solution cannot be found at the current search step.
As a result, there are at most $\lceil \log N \rceil$ iterations, where $N$ is the number of workers. 
In each iteration, the linear programming can be solved in polynomial time according to \cite{spielman2004smoothed}.
Actually, since the base topology in real world is usually sparse, the practical time cost for Alg. \ref{alg:algorithm} will be further reduced at the coordinator, which is usually deployed in cloud or cloudlet with high computing power.
Therefore, the time for solving the joint optimization problem can be negligible, compared with that for model training and transmission.

% \vspace{-0.7em}
\section{Experimentation and Evaluation}\label{sec:results}
% This section first describes the datasets and models for experiments (Sec. \ref{dataset}).
% Then we introduce baselines and metrics for performance comparison (Sec. \ref{baselines}).
% Finally, experimental settings and results are presented (Sec. \ref{simulation}). 

\subsection{Datasets and Models}\label{dataset}
\textbf{Datasets:}
We conduct extensive experiments on three real-world datasets: (\romannumeral1) EMNIST, (\romannumeral2) CIFAR-10, and (\romannumeral3) ImageNet.
Specifically, EMNIST \cite{cohen2017emnist} is a handwritten character dataset that contains 731,668 training samples and 82,587 test samples from 62 categories (10 digits, 52 characters with lowercase and uppercase).
%CIFAR-10 contains 60,000 32$\times$32 color images labeled in 10 classes with 50,000 samples for training and 10,000 samples for test.
CIFAR-10 is an image dataset composed of 60,000 32$\times$32 colour images (50,000 for training and 10,000 for test) in 10 categories.
ImageNet \cite{russakovsky2015imagenet} is a dataset for visual recognition which consists of 1,281,167 training images, 50,000 validation images and 100,000 test images from 1,000 categories.
%  and the dimension of each image is 224$\times$224$\times$3.
% each of which is a 224*224 color image.
% , and each image from ImageNet is a 224$\times$224 color image.
To cope with the constrained resource of edge devices, we create IMAGE-100, a subset of ImageNet that contains 100 out of 1,000 categories, and each sample is resized with the shape of 64$\times$64$\times$3.

To simulate the non-IID setting, we propose to create synthesized non-IID datasets with different \textit{class distribution skews} as in \cite{zhao2018federated,wang2020optimizing}, \eg, a single user can possess more data for one class or a couple of classes than others.
Concretely, $p$ (\eg, 0.1, 0.2, 0.4, 0.6 and 0.8) of a unique class is divided equally for every three workers and the remaining samples of each class are partitioned to other workers uniformly.
Accordingly, the non-IID levels of the above datasets are denoted as 0.1, 0.2, 0.4, 0.6 and 0.8, respectively.
Note that $p$ = 0.1 is a special case, where the distribution of training dataset is IID for 30 workers.
For fair comparisons, the full test datasets are used across all workers.
% For fair comparison, the test datasets are partitioned uniformly among workers.

\textbf{Models:} Three models with different types and structures are implemented on the above three real-world datasets for performance evaluation:
(\romannumeral1) CNN on EMNIST, (\romannumeral2) AlexNet on CIFAR-10, (\romannumeral3) VGG-16 on IMAGE-100.
Firstly, The plain CNN model \cite{mcmahan2017communication} specialized for the EMNIST dataset has two 5$\times$5 convolutional layers, a fully-connected layer with 512 units, and a softmax output layer with 62 units.
Secondly, An 8-layer AlexNet \cite{krizhevsky2012imagenet}, which is composed of three 3$\times$3 convolutional layers, one 7$\times$7 convolutional layer, one 11$\times$11 convolutional layer, two fully-connected hidden layers, and one fully-connected output layer, is adopted for CIFAR-10. 
Thirdly, a famous model VGG-16 \cite{simonyan2014very}, that consists of 13 convolution layers with kernel of 3$\times$3, two dense layers and a softmax output layer, is utilized to classify the images in IMAGE-100.

\subsection{Baselines and Metrics}\label{baselines}
\textbf{Baselines:} We choose four classical algorithms as baselines for performance comparison, which are summarized as follows.
(\romannumeral1) D-PSGD \cite{lian2017can} is a synchronous DFL algorithm using a ring network topology and the same local updating frequency for workers.
(\romannumeral2) AD-PSGD \cite{lian2018asynchronous} is an asynchronous DFL algorithm, where workers randomly send local models to one of their neighbors immediately after performing local updating to speed up the training process.
(\romannumeral3) LD-SGD \cite{li2019communication} alternates the frequencies of local updating and global updating for efficient decentralized communication.
(\romannumeral4) PENS \cite{onoszko2021decentralized} with adaptive network topology allows workers with similar data distributions to communicate with each other to deal with statistical heterogeneity.
 
\textbf{Metrics:} The following metrics are adopted to evaluate the performance of FedHP and the baselines.
(\romannumeral1) \textit{Test accuracy} is measured by the proportion between the amount of the right data predicted by the model and that of all data. Specifically, at each communication round, we evaluate the average test accuracy of all workers' models trained with different algorithms on the test datasets.
(\romannumeral2) \textit{Completion time} is defined as the total training time until the average model of all workers converges to the target accuracy. 
Concretely, we record the completion time of each communication round and sum up to get the total training time.
(\romannumeral3) \textit{Average waiting time} is introduced to reflect the training efficiency of different algorithms.
Specifically, the waiting time of worker $i$ at round $h$ can be represented by $t^h-t_i^h$, then the average waiting time of all workers at round $h$ is expressed as $\frac{1}{N} \sum_{i=1}^{N}(t^h-t_i^h)$.

\subsection{Experiments}\label{simulation}
\subsubsection{Experimental Setup}
We evaluate the performance of FedHP through extensive simulation experiments, which are conducted on an AMAX deep learning workstation equipped with an Intel(R) Xeon(R) Gold 5218R CPU, 8 NVIDIA GeForce RTX 3090 GPUs and 256 GB RAM.
On the workstation, we simulate a heterogeneous EC system with 30 workers and one coordinator (each is implemented as a process in the system) for DFL.
The implementation for model training on each worker is based on the PyTorch framework \cite{paszke2019pytorch}, and we use the socket library of Python to build up the communication among workers and between workers and the coordinator.

%To simulate the heterogeneous communication capacities of workers,
We consider the common situation where each worker communicates with its neighbors and coordinator through either LANs or WANs.
To reflect the heterogeneity and dynamics of networks in our simulations, we let the bandwidth of each worker fluctuate between 1Mb/s and 10Mb/s.
% Considering the outbound bandwidth in typical WANs is usually smaller than the inbound bandwidth \cite{mcmahan2017communication}, we configure it to fluctuate between 0.5Mb/s and 5Mb/s.
In addition, for simulating the computing heterogeneity, we assume that the computing time of one local iteration on a certain simulated worker is subject to the Gaussian distribution. 
Different simulated workers are randomly assigned with a specific Gaussian function whose mean and variance are derived from the time records of performing one local iteration on a commercial device (\eg, laptop, Jetson TX, Xavier NX).

Each experiment will by default run 200, 500, and 500 communication rounds for EMNIST, CIFAR-10 and IMAGE-100, respectively, which will guarantee the convergence of the models.
For CNN on EMNIST, the learning rate is initialized as 0.1 and the corresponding decay rate is specified as 0.98, while for AlexNet on CIFAR-10 and VGG-16 on IMAGE-100, the learning rates and the corresponding decay rates of them are identical, separately initialized as 0.1 and 0.993 \cite{xu2022adaptive}.
Besides, the batch size is set as 32 for all three models.

\begin{figure}[t]
	\centering
	%		\vspace{5pt}
	\subfigure[EMNIST]
	{
		\includegraphics[width=0.29\linewidth,height=2.3cm]{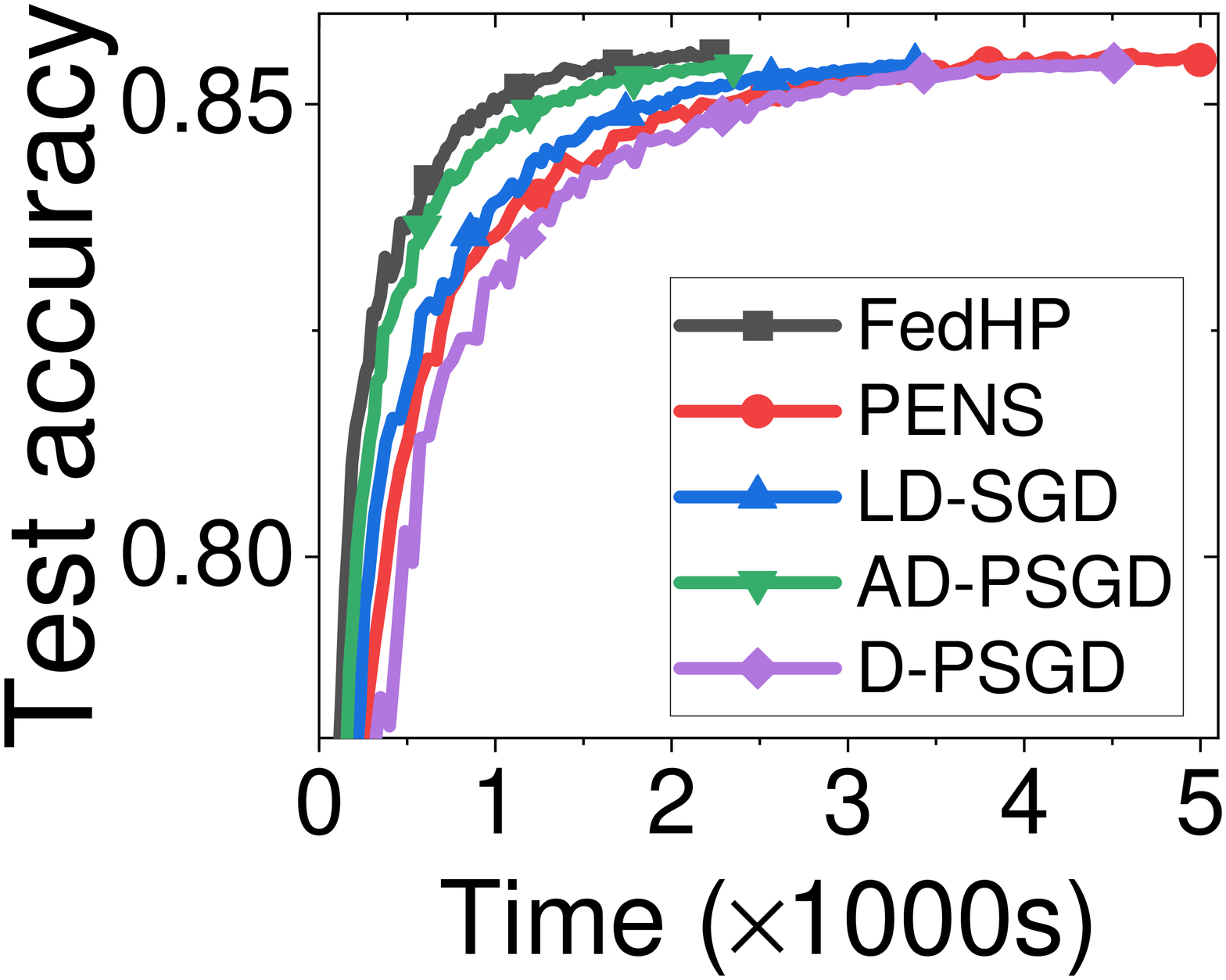}
		\label{fig:EMNIST-IID}
	}
	\subfigure[CIFAR-10]
	{
		\includegraphics[width=0.29\linewidth,height=2.3cm]{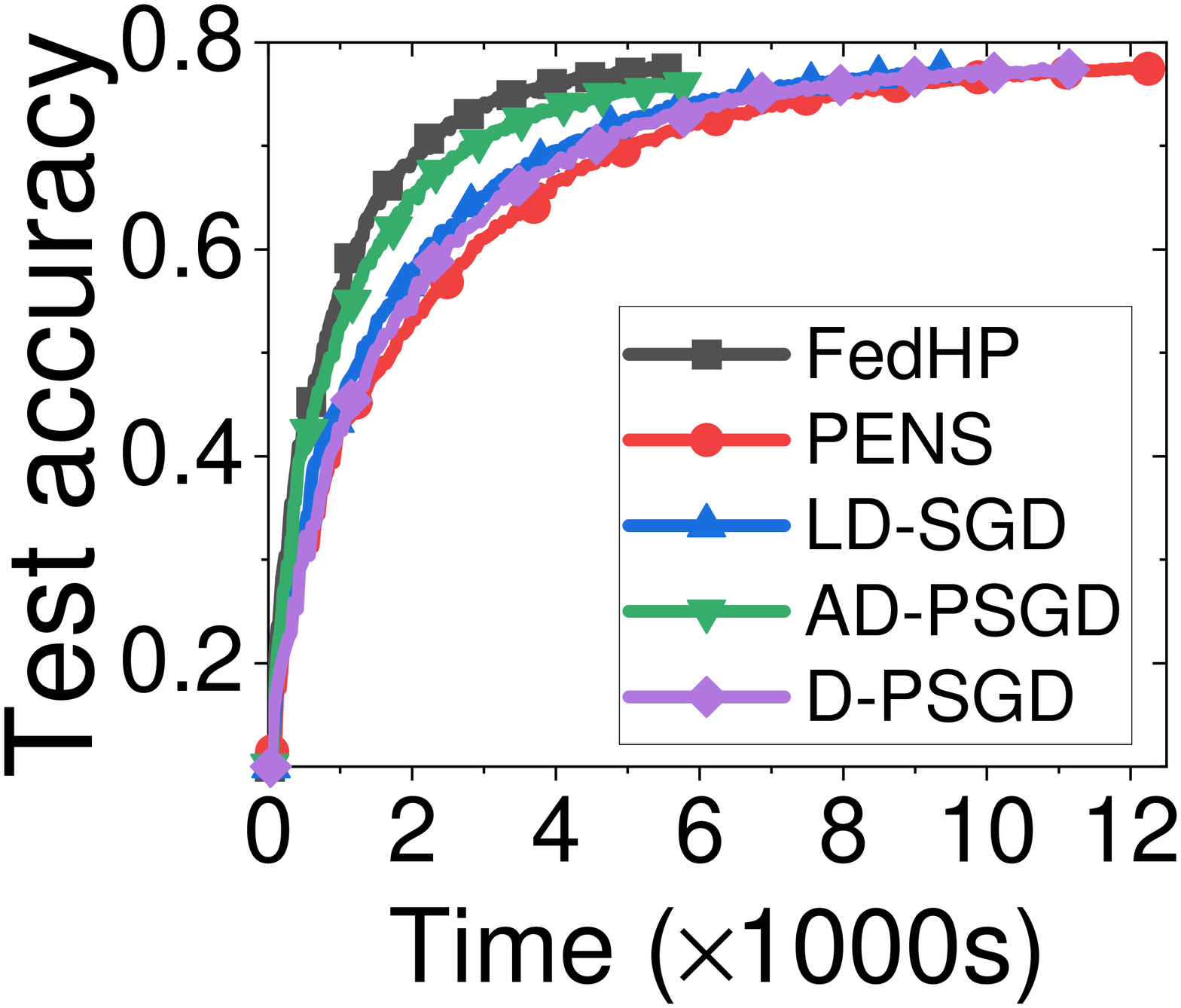}
		\label{fig:CIFAR10-IID}
	}
	\subfigure[IMAGE-100]
	{
		\includegraphics[width=0.29\linewidth,height=2.3cm]{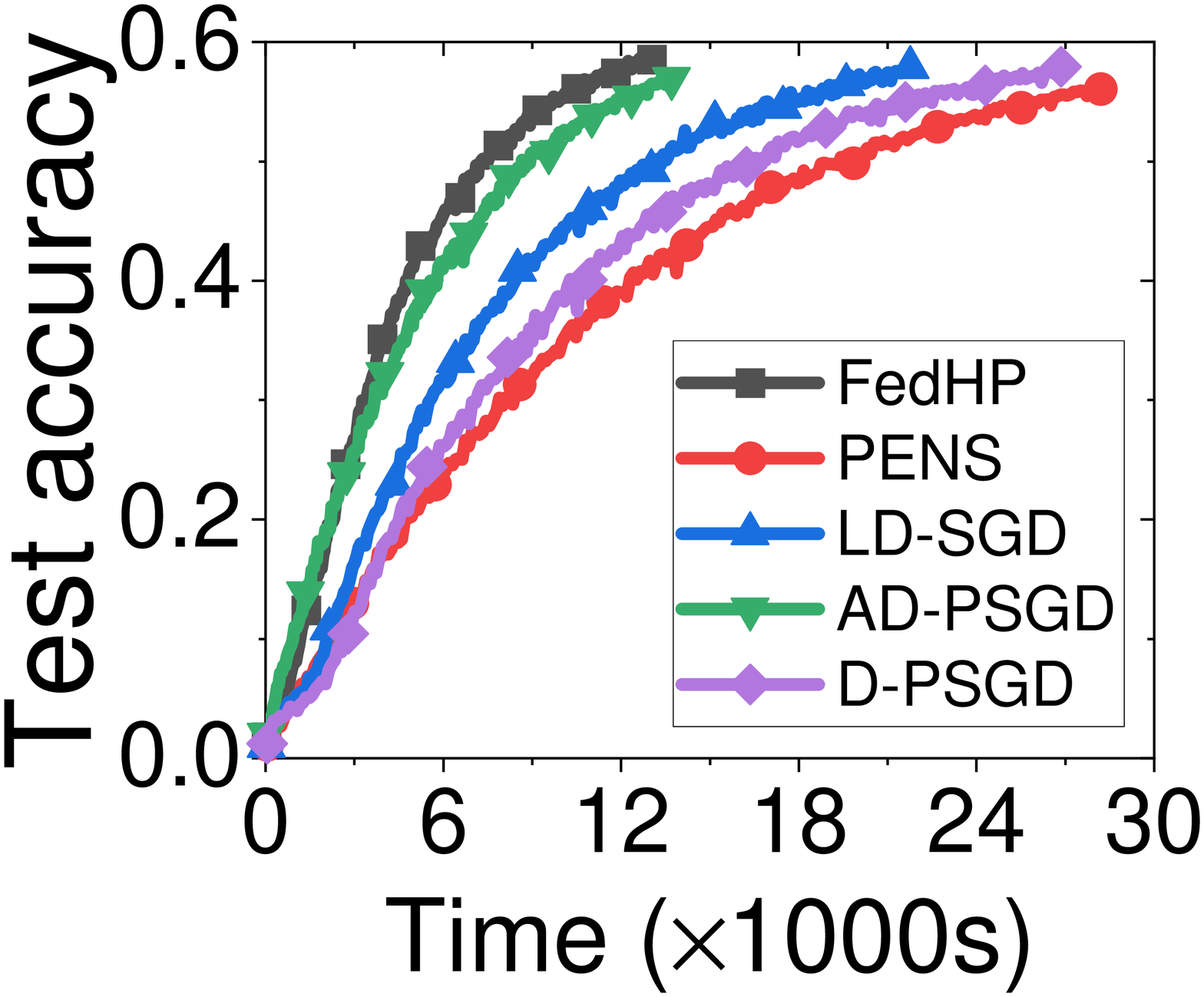}
		\label{fig:IMAGE100-IID}
	}
	\caption{Test accuracy of five algorithms on the three IID datasets.}
	\label{fig:IID}
	\vspace{-0.6em}
\end{figure}

\begin{figure}[t]
	\centering
	%		\vspace{5pt}
	\subfigure[EMNIST]
	{
		\includegraphics[width=0.29\linewidth,height=2.3cm]{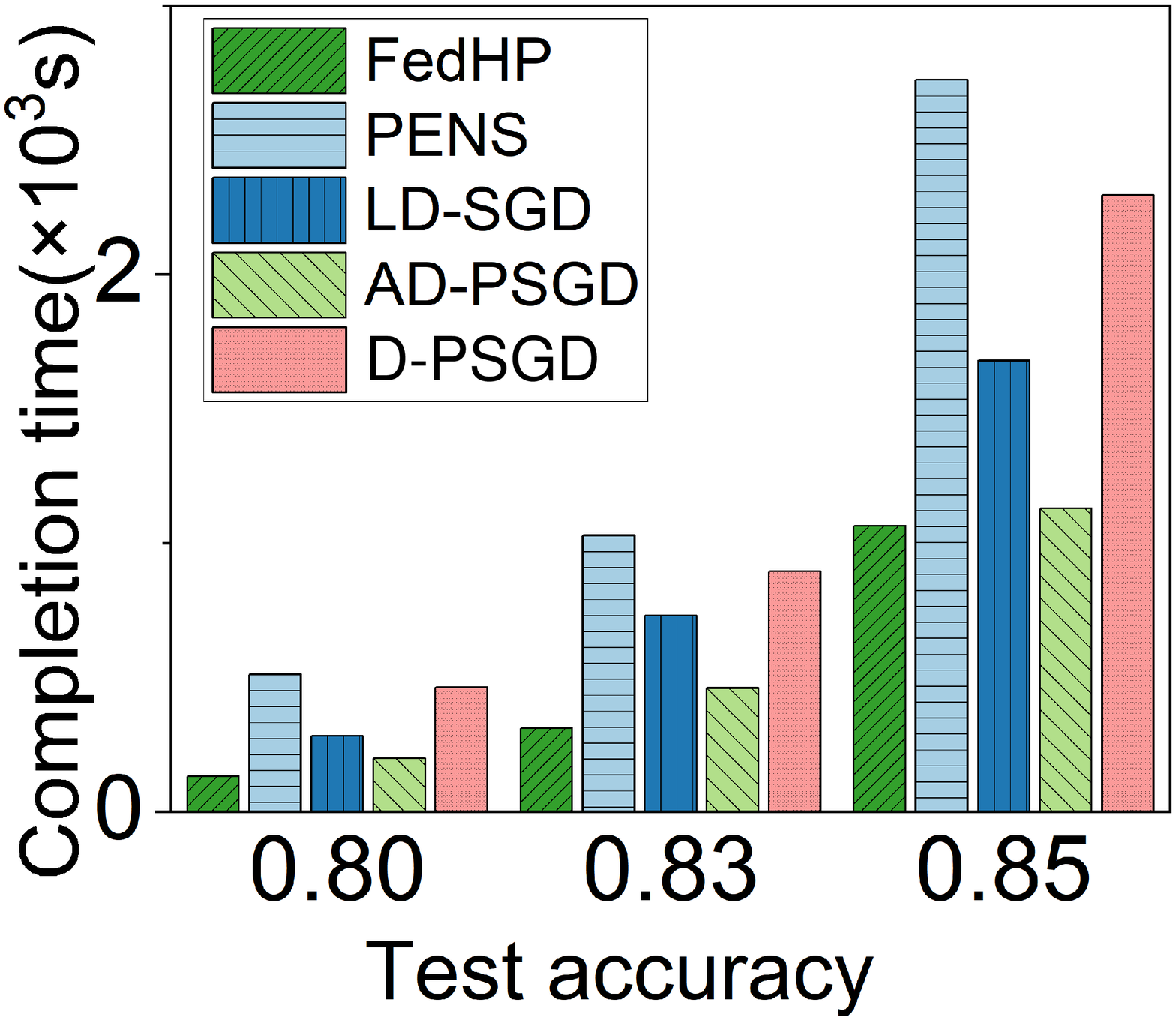}
		\label{fig:EMNIST_time}
	}
	\subfigure[CIFAR-10]
	{
		\includegraphics[width=0.29\linewidth,height=2.3cm]{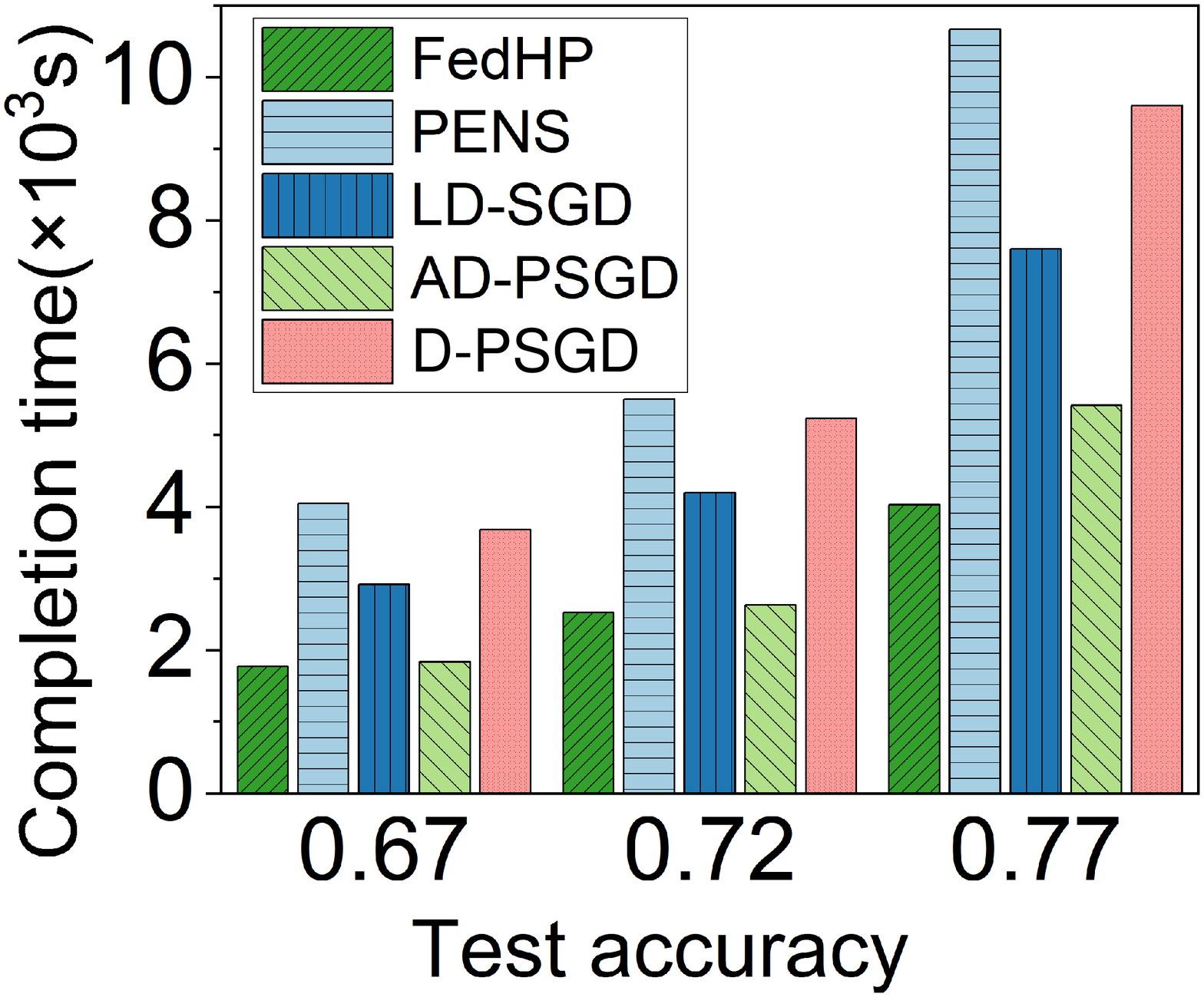}
		\label{fig:CIFAR10_time}
	}
	\subfigure[IMAGE-100]
	{
		\includegraphics[width=0.29\linewidth,height=2.3cm]{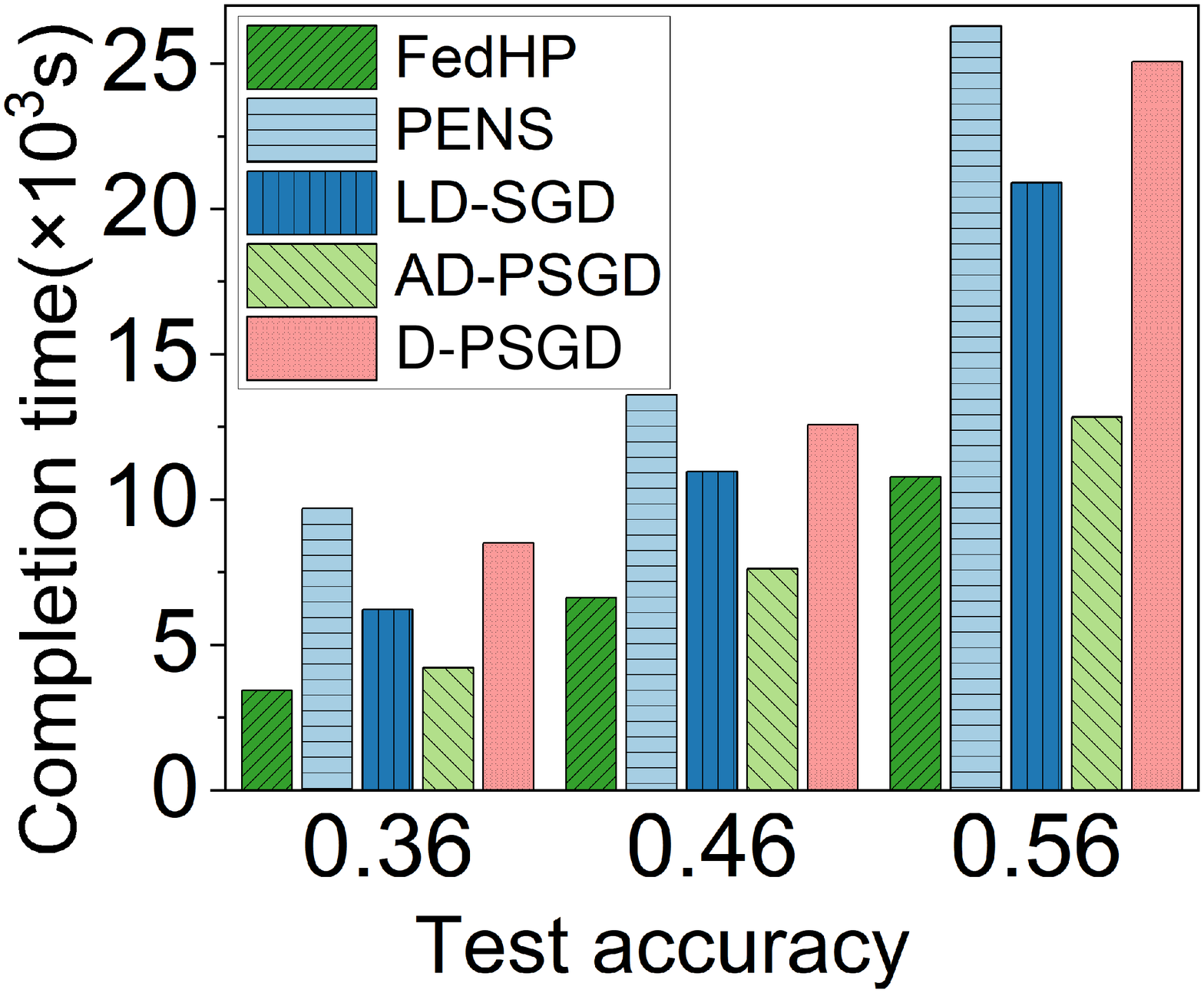}
		\label{fig:IMAGE100_time}
	}
	\caption{Completion time of five algorithms when achieving different target accuracy}
	\label{fig:completion_time}
	\vspace{-0.9em}
\end{figure}

\begin{figure}[t]
	\centering
	%		\vspace{5pt}
	\subfigure[EMNIST]
	{
		\includegraphics[width=0.29\linewidth,height=2.3cm]{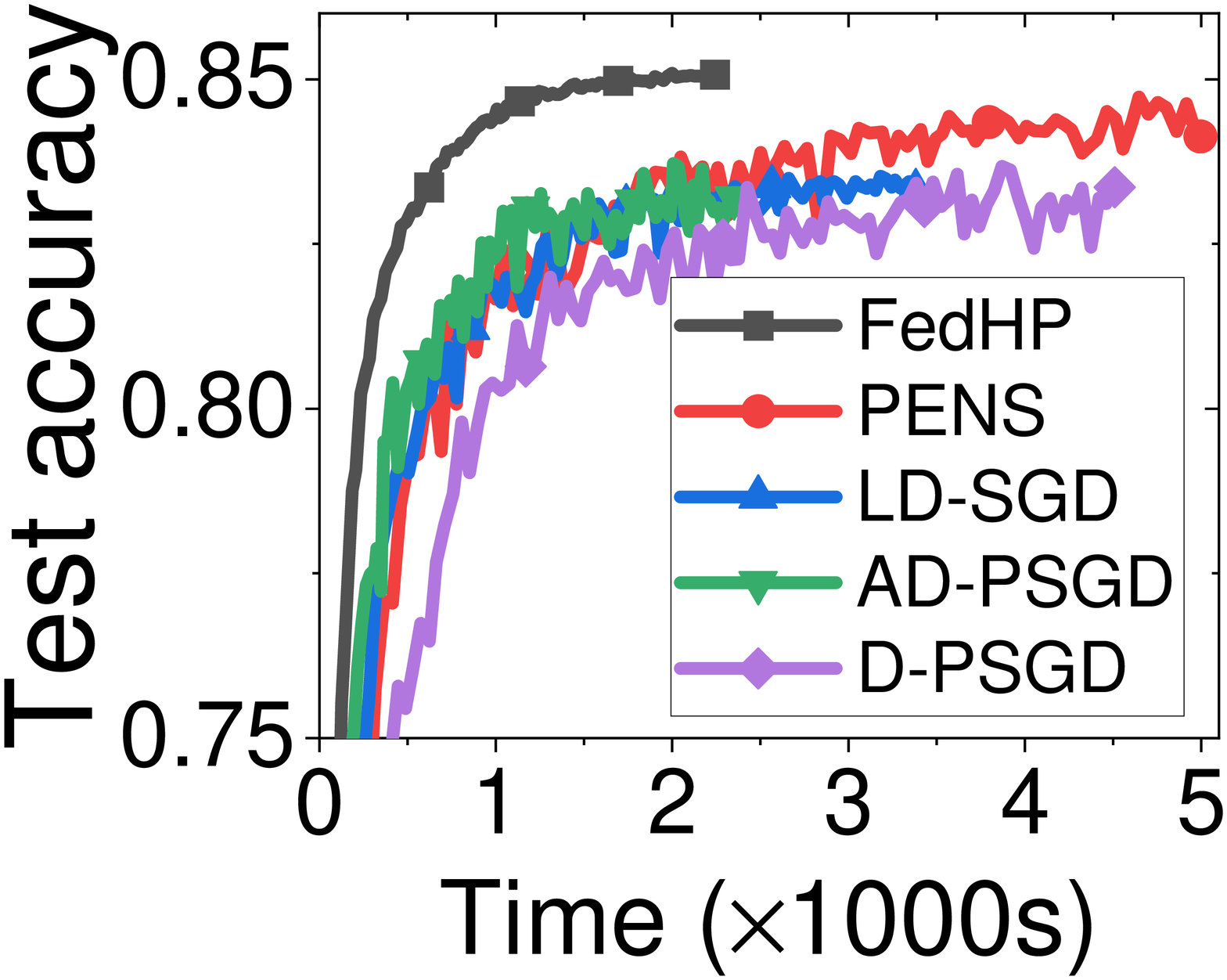}
		\label{fig:EMNIST-non-IID0.6}
	}
	\subfigure[CIFAR-10]
	{
		\includegraphics[width=0.29\linewidth,height=2.3cm]{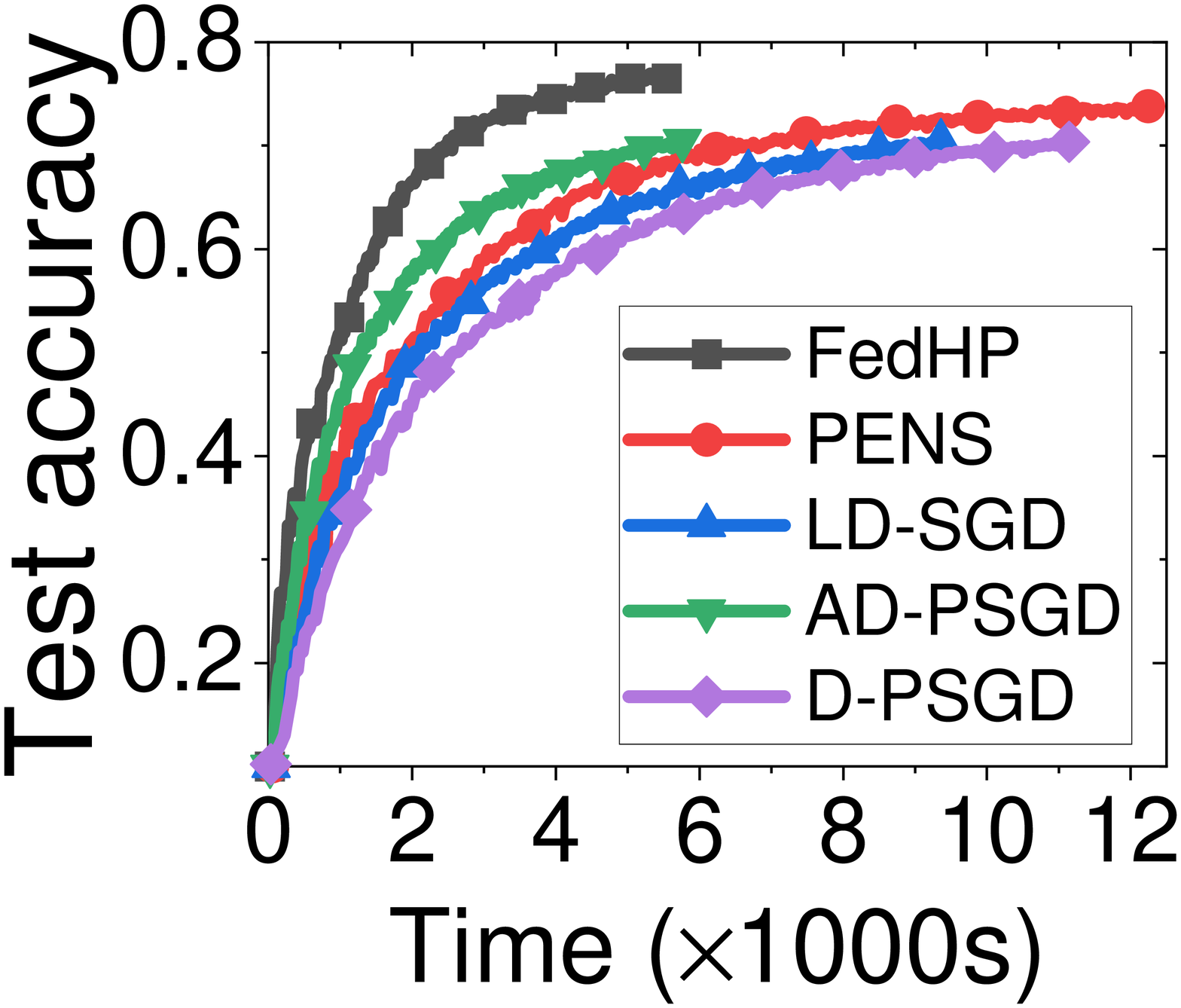}
		\label{fig:CIFAR10-non-IID0.6}
	}
	\subfigure[IMAGE-100]
	{
		\includegraphics[width=0.29\linewidth,height=2.3cm]{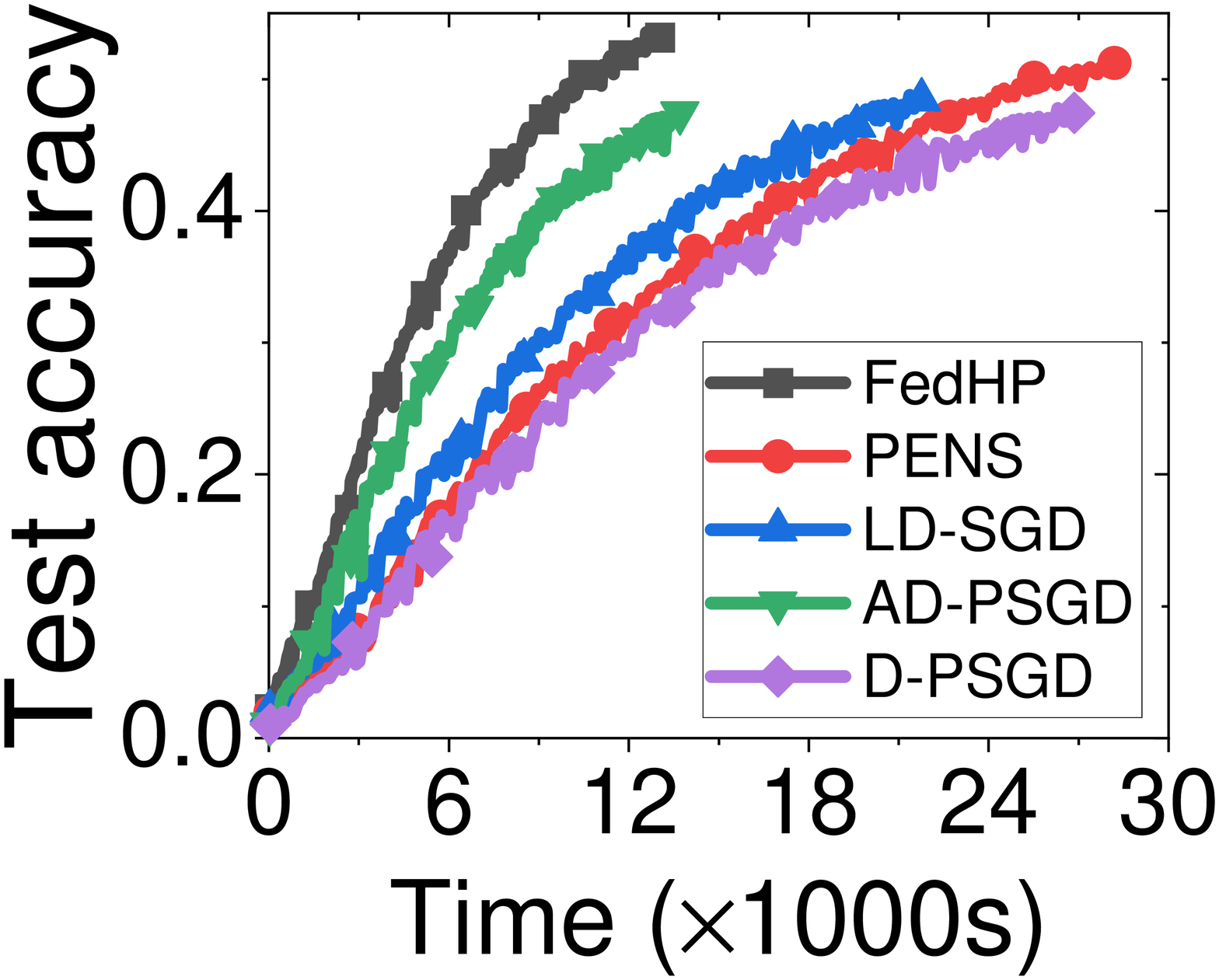}
		\label{fig:IMAGE100-non-IID0.6}
	}
	\caption{Test accuracy of five algorithms on the three datasets with non-IID level $p$=0.6.}
	\label{fig:non-IID0.6}
	\vspace{-0.9em}
\end{figure}

\begin{figure}[t]
	\centering
	%		\vspace{5pt}
	\subfigure[EMNIST]
	{
		\includegraphics[width=0.29\linewidth,height=2.3cm]{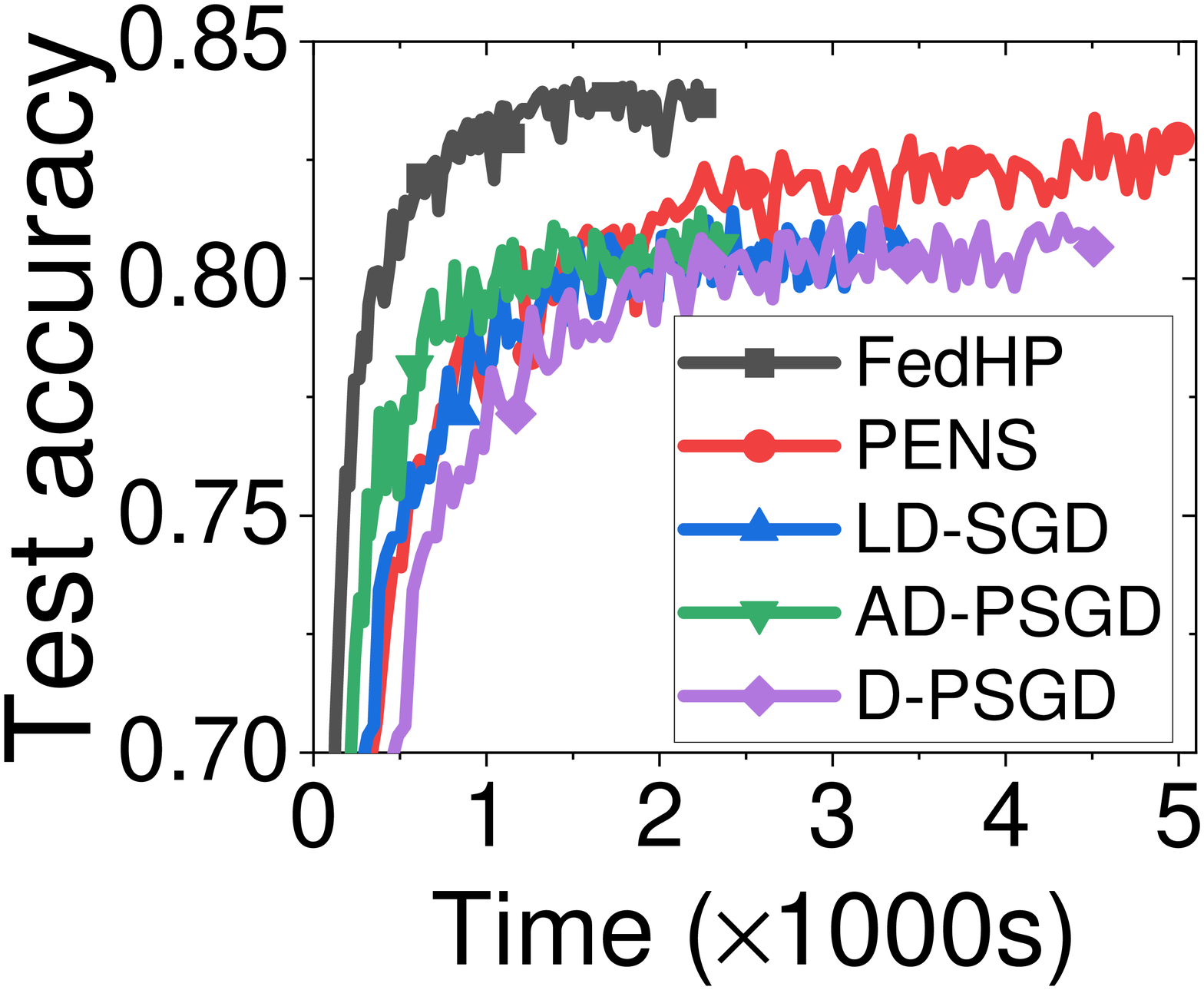}
		\label{fig:EMNIST-non-IID0.8}
	}
	\subfigure[CIFAR-10]
	{
		\includegraphics[width=0.29\linewidth,height=2.3cm]{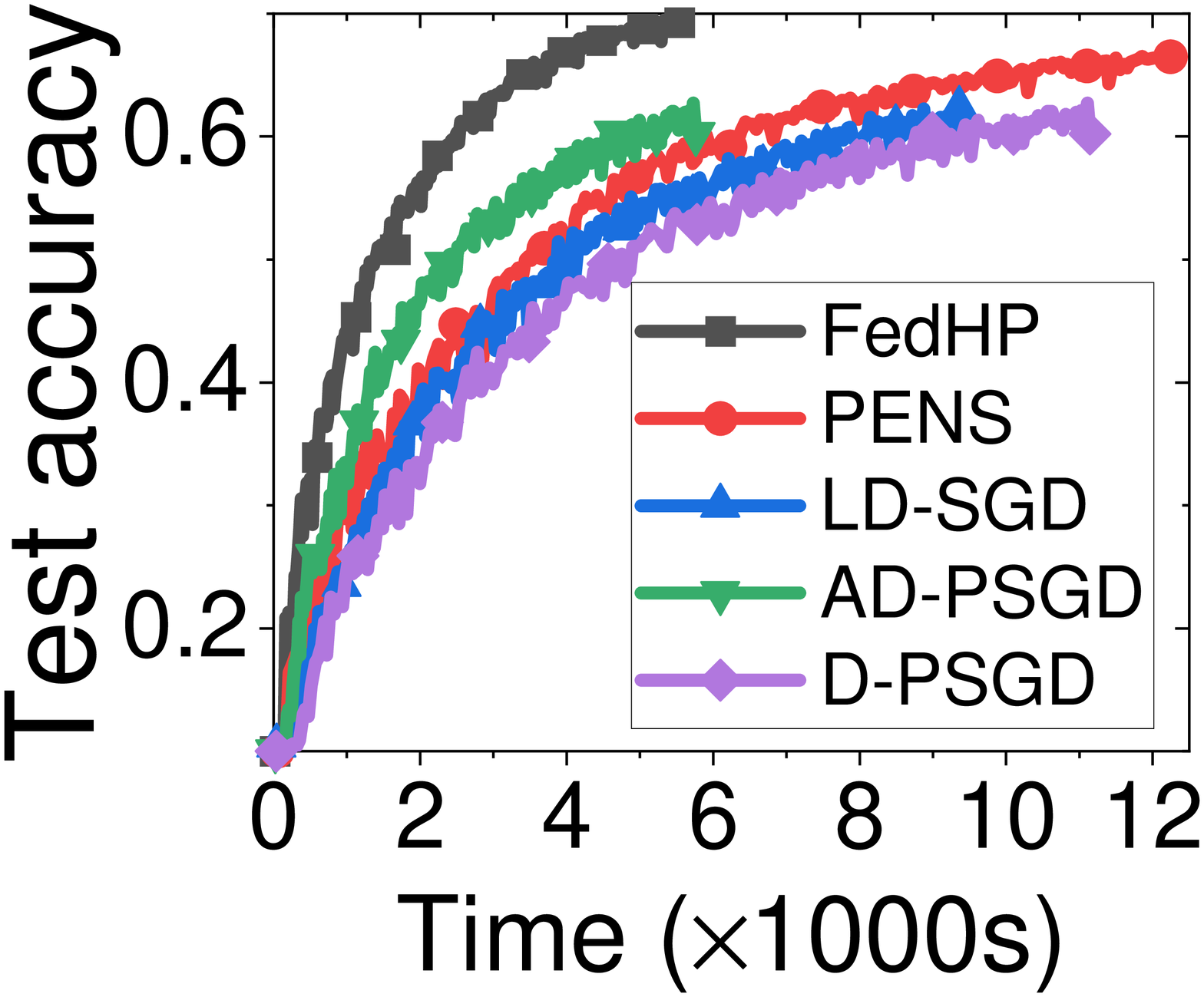}
		\label{fig:CIFAR10-non-IID0.8}
	}
	\subfigure[IMAGE-100]
	{
		\includegraphics[width=0.29\linewidth,height=2.3cm]{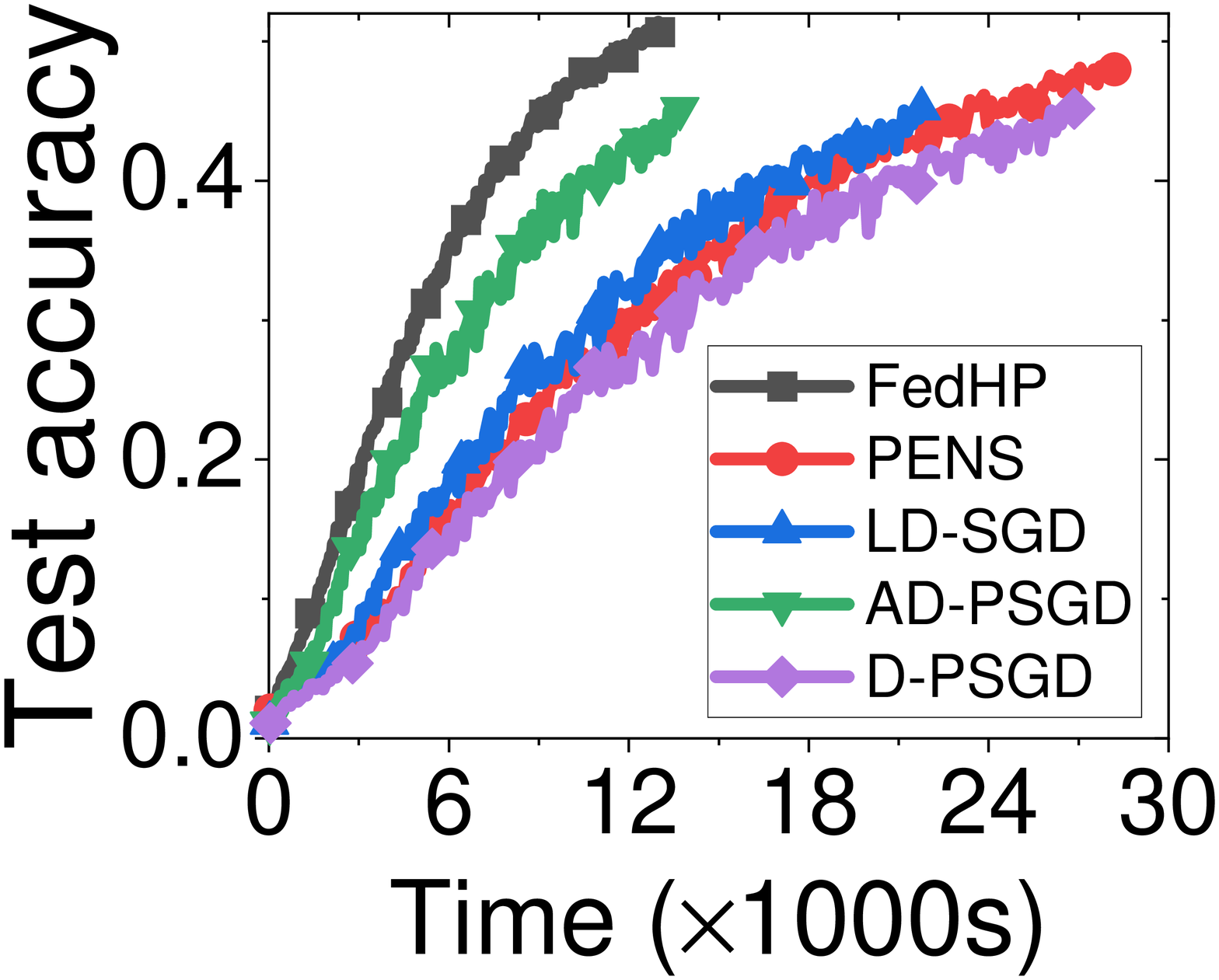}
		\label{fig:IMAGE100-non-IID0.8}
	}
	\caption{Test accuracy of five algorithms on the three datasets with non-IID level $p$=0.8.}
	\label{fig:non-IID0.8}
	\vspace{-0.9em}
\end{figure}

\begin{figure}[t]
	\centering
	%		\vspace{5pt}
	\subfigure[EMNIST]
	{
		\includegraphics[width=0.29\linewidth,height=2.3cm]{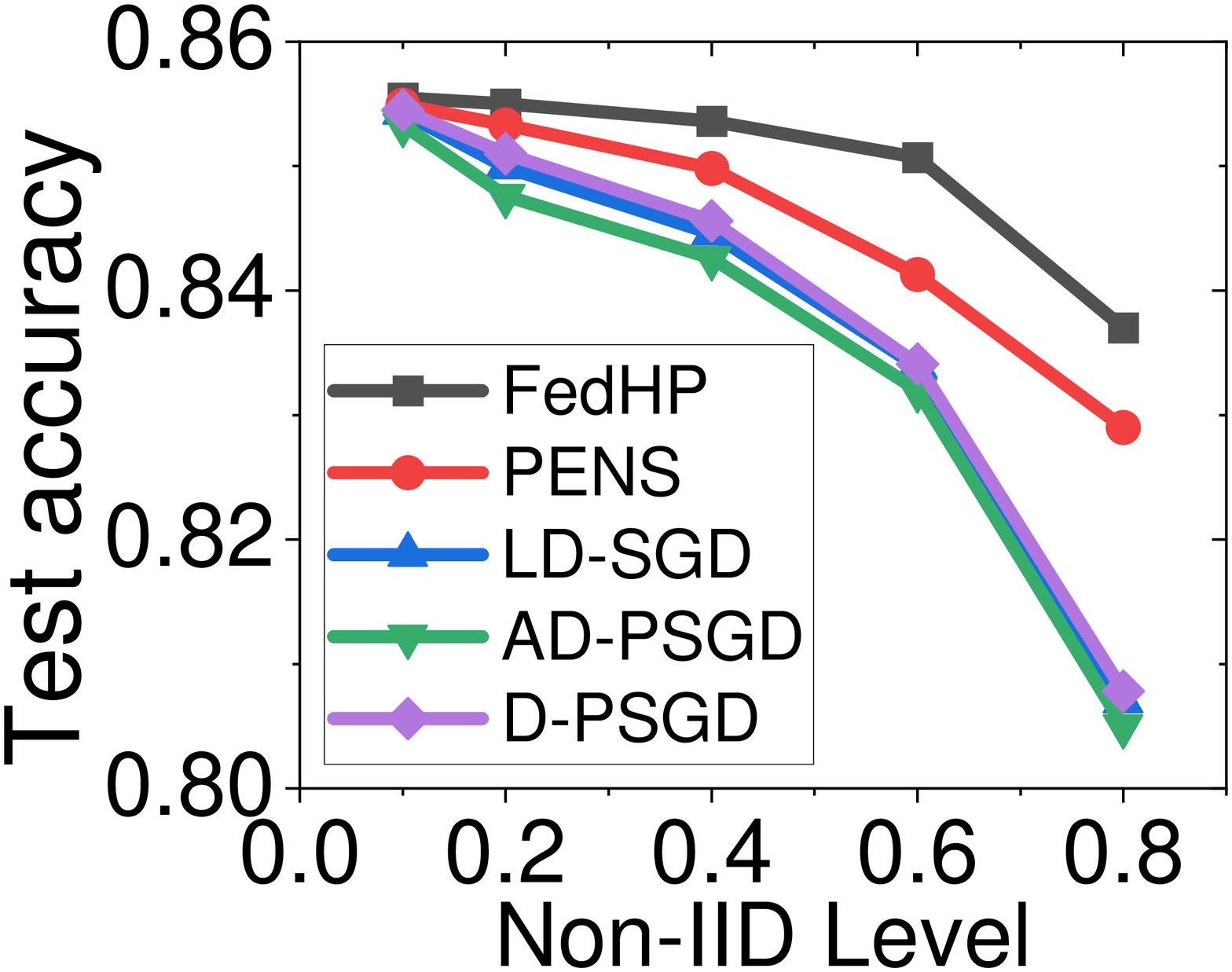}
		\label{fig:EMNIST-non-IID_level}
	}
	\subfigure[CIFAR-10]
	{
		\includegraphics[width=0.29\linewidth,height=2.3cm]{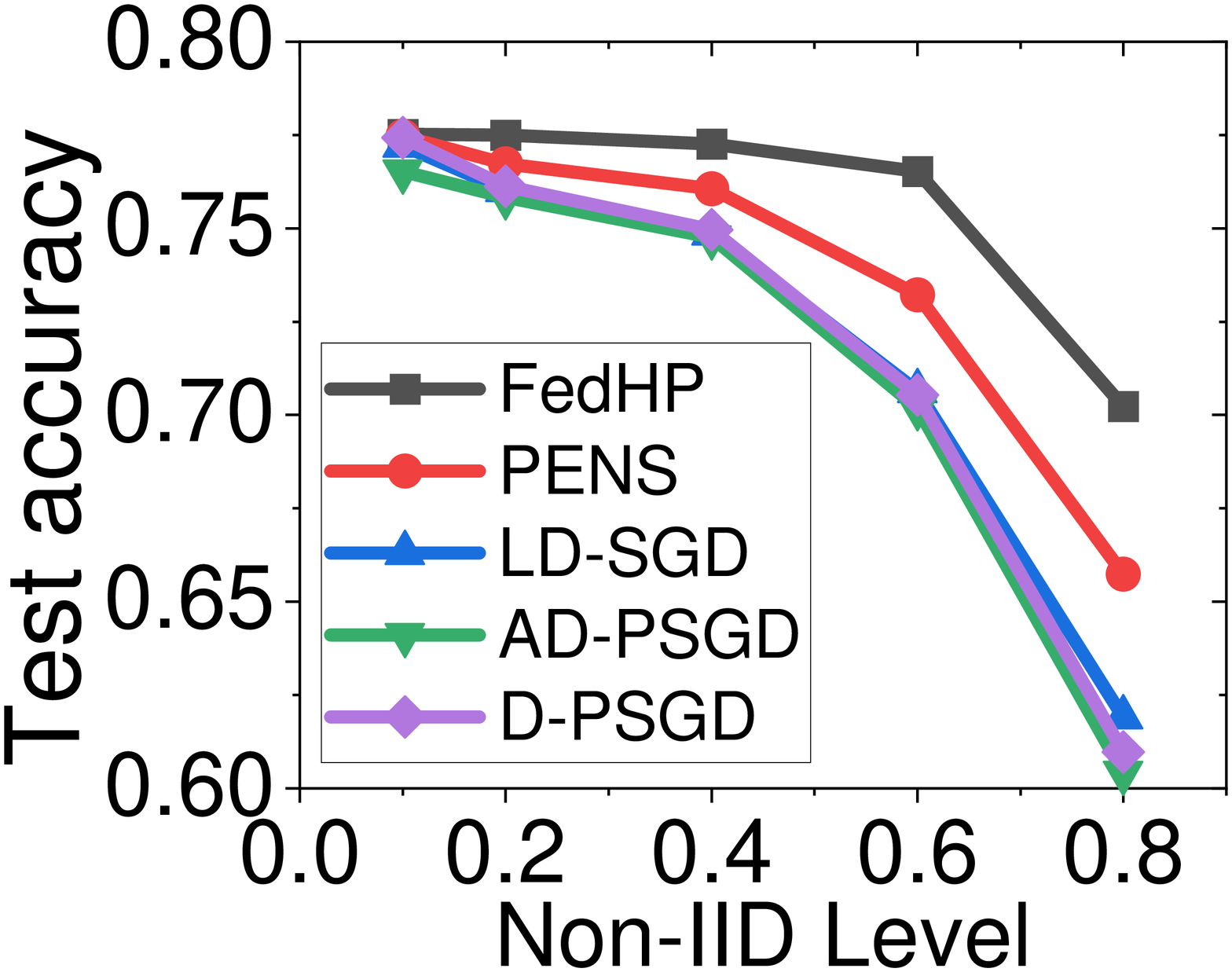}
		%\label{fig:C10bandwidth}
	}
	\subfigure[IMAGE-100]
	{
		\includegraphics[width=0.29\linewidth,height=2.3cm]{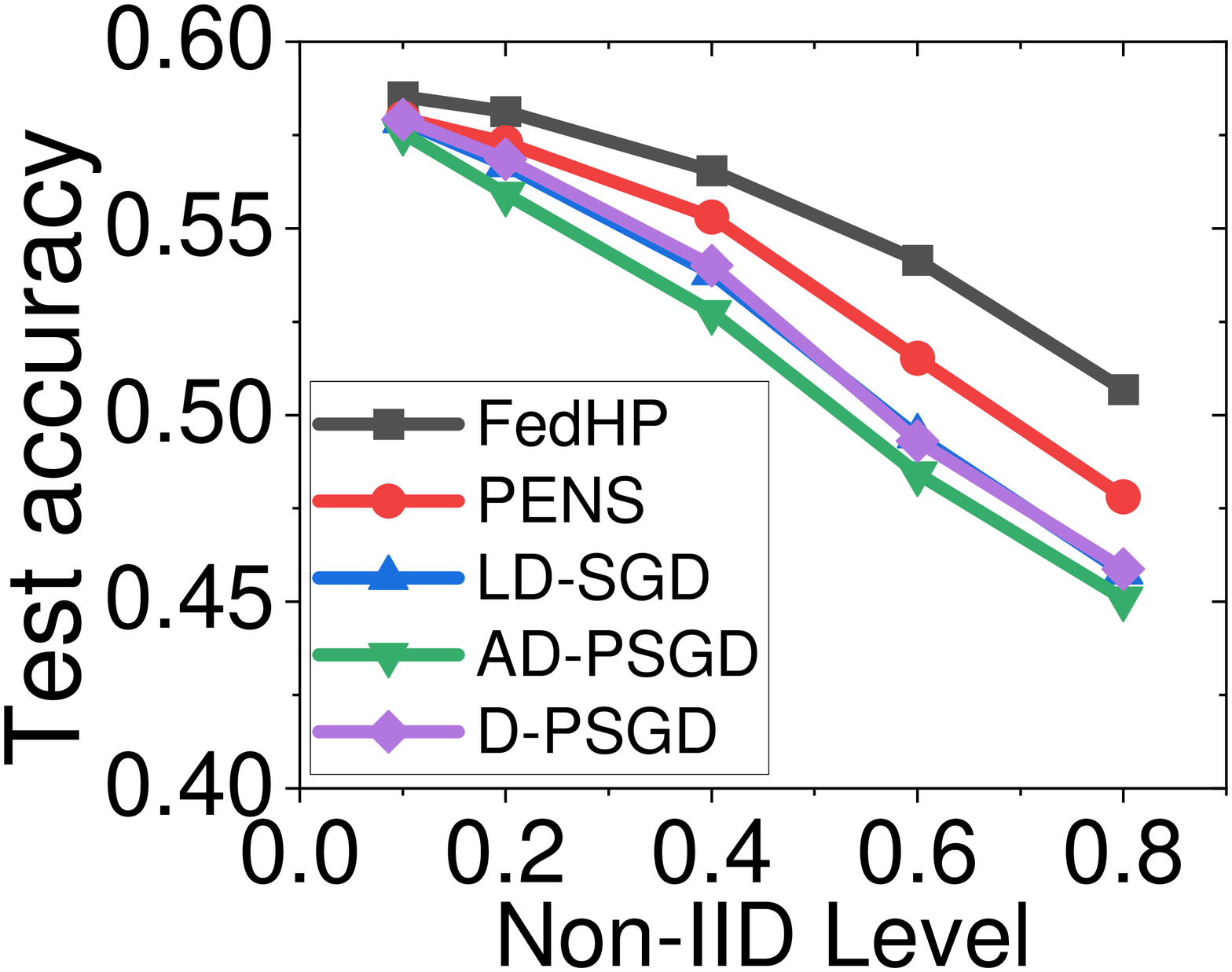}
		\label{fig:IMAGE100-non-IID_level}
	}
	\caption{Test accuracy varies with different non-IID levels.}
	\label{fig:non-IID_level}
	\vspace{-0.9em}
\end{figure}

\begin{figure}[t]
	\centering
	%		\vspace{5pt}
	\subfigure[EMNIST]
	{
		\includegraphics[width=0.29\linewidth,height=2.3cm]{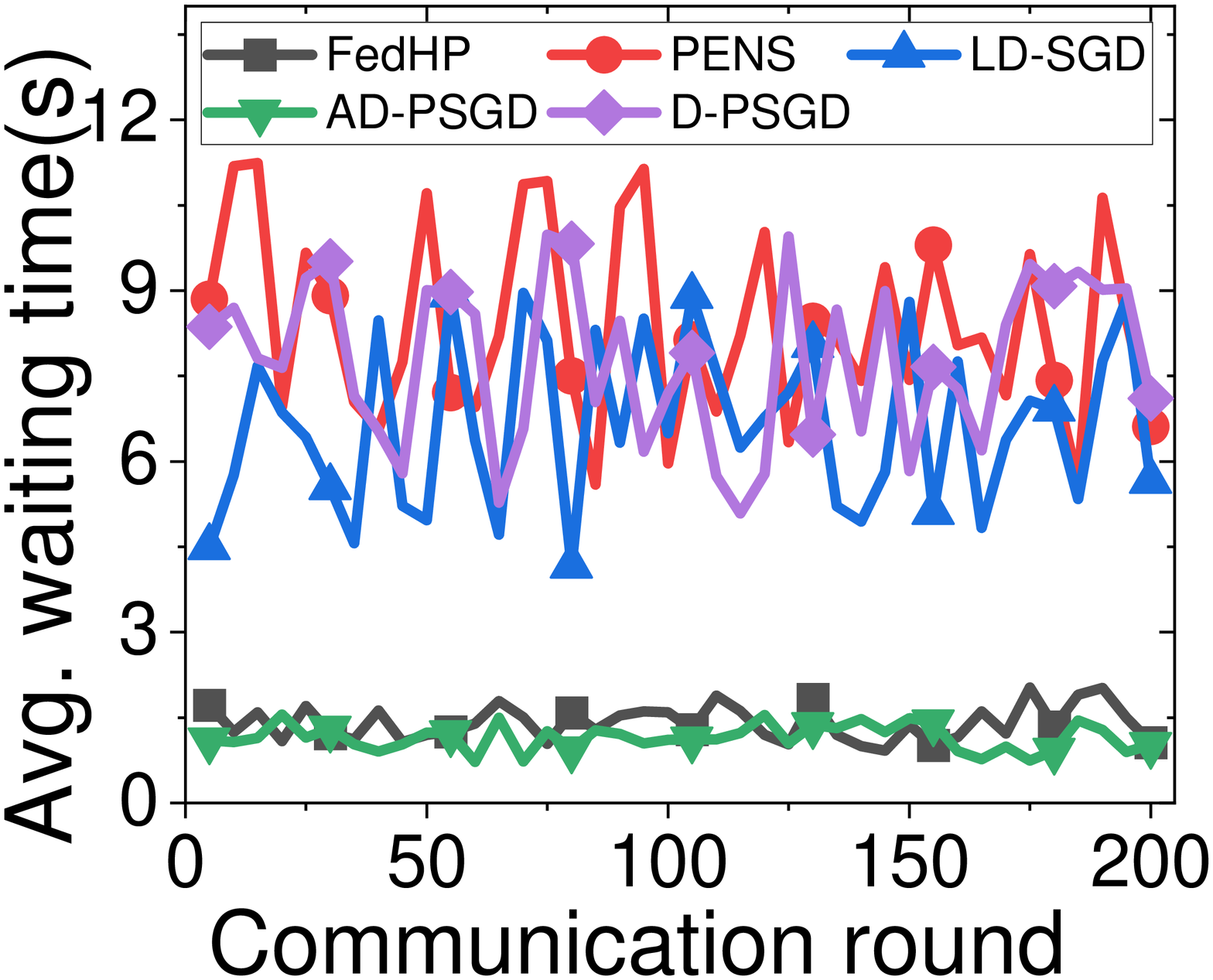}
	}
	\subfigure[CIFAR-10]
	{
		\includegraphics[width=0.29\linewidth,height=2.3cm]{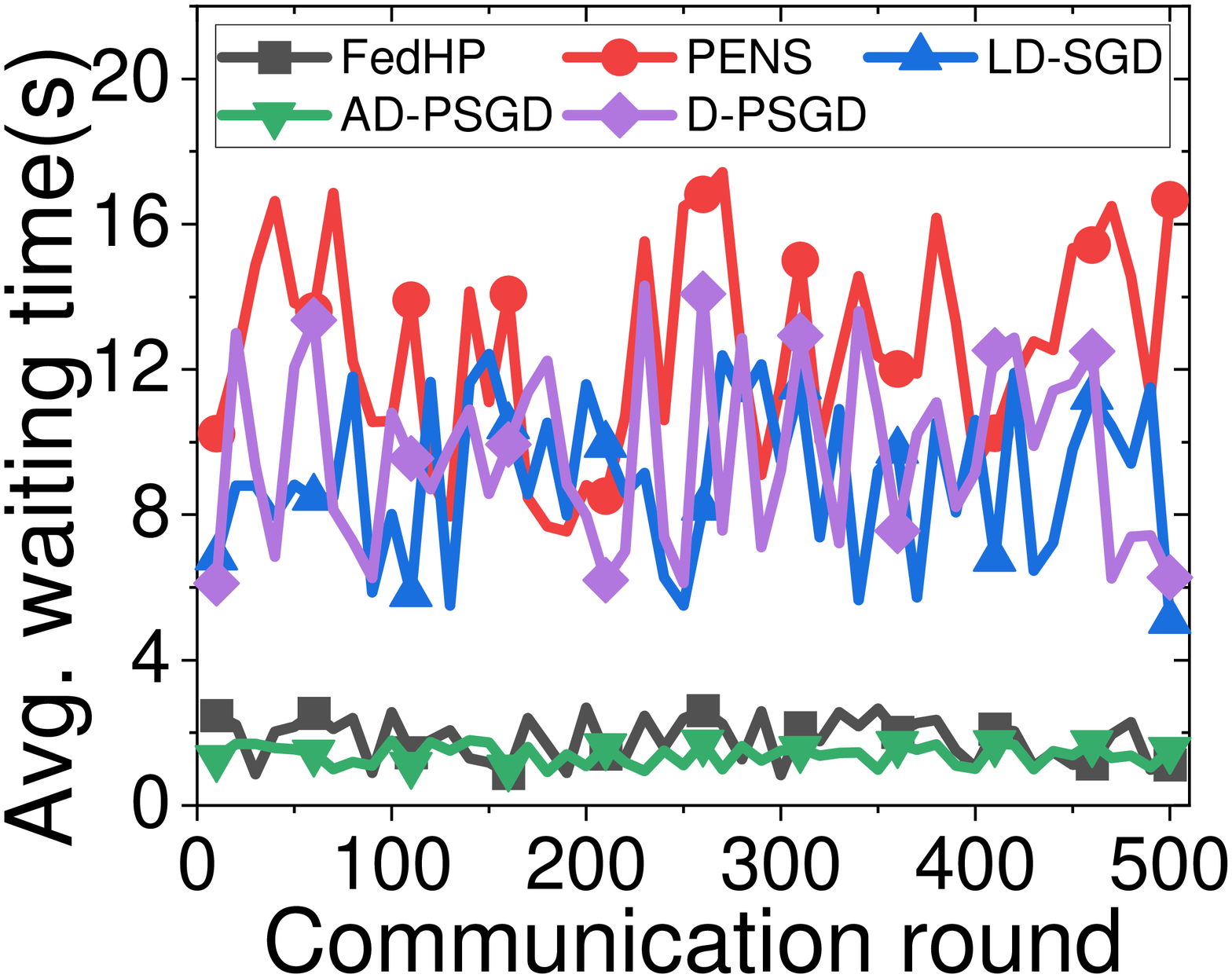}
		\label{fig:CIFAR10-waiting_time}
	}
	\subfigure[IMAGE-100]
	{
		\includegraphics[width=0.29\linewidth,height=2.3cm]{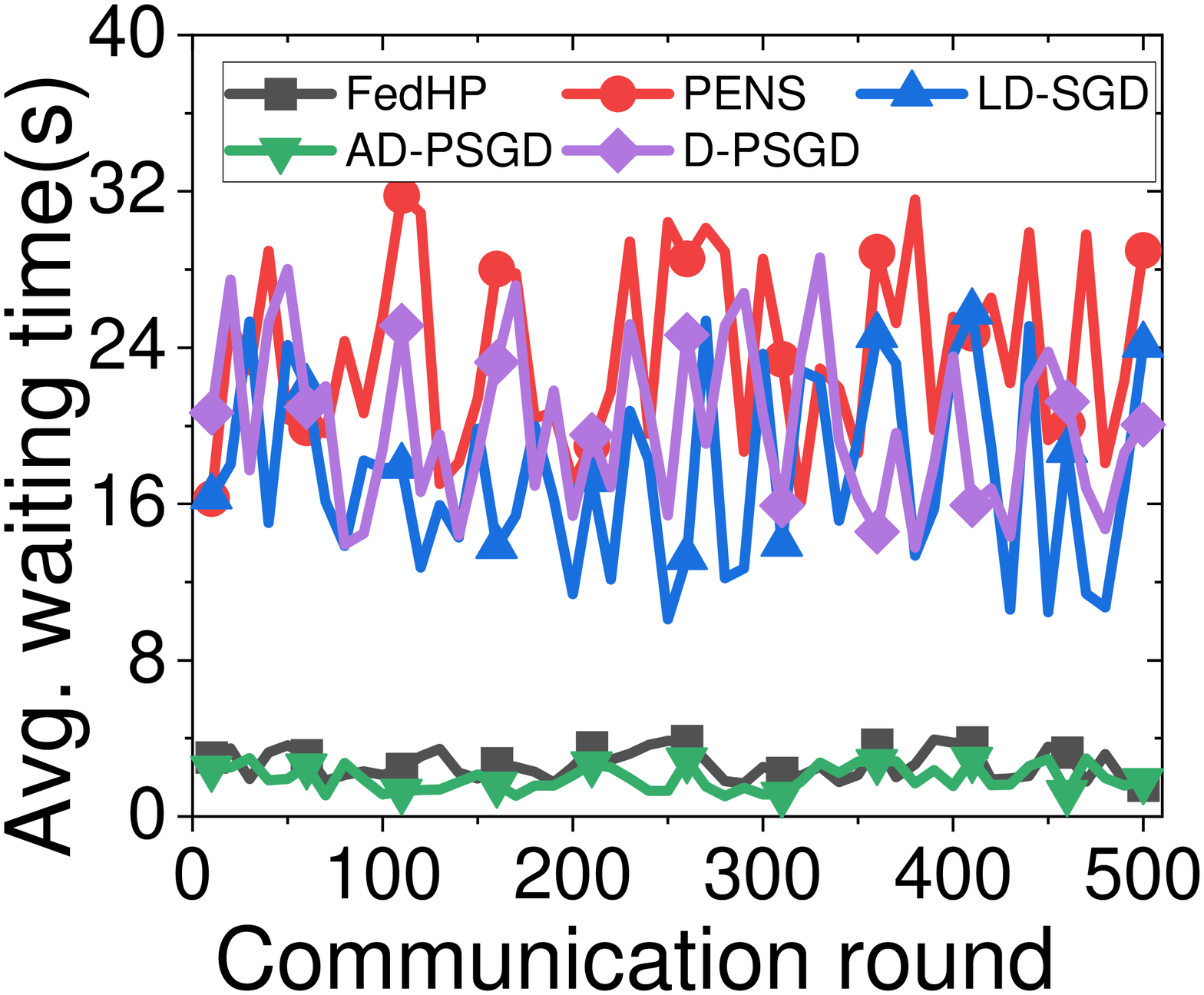}
		\label{fig:ImageNet-waiting_time}
	}
	\caption{Average waiting time of five algorithms on the three datasets.}
	\label{fig:waiting_time}
	\vspace{-0.9em}
\end{figure}

\subsubsection{Overall Effectiveness}
Firstly, we implement a set of experiments of these algorithms on the IID datasets.
The training processes of FedHP and the baselines are presented in Fig. \ref{fig:IID}.
In addition, we show the completion time of different algorithms when they achieve different target accuracy in Fig. \ref{fig:completion_time}.
The results demonstrate that all the algorithms achieve the similar test accuracy eventually.
FedHP achieves the fastest convergence, followed by AD-PSGD on all the three datasets, and they are much faster than the other methods.
For example, by Figs. \ref{fig:EMNIST-IID} and \ref{fig:EMNIST_time}, FedHP takes 1,064s to achieve 85\% accuracy for CNN on EMNIST, while PENS, LD-SGD, AD-PSGD, D-PSGD, takes 2,725s, 1,680s, 1,129s, 2,254s, respectively.
Besides, by Figs. \ref{fig:CIFAR10-IID} and \ref{fig:CIFAR10_time}, FedHP reduces the completion time of training AlexNet by about 56\%, 41\%, 3\% and 51\%, compared with PENS, LD-SGD, AD-PSGD and D-PSGD.
Moreover, for VGG-16 on IMAGE-100 as shown in Figs. \ref{fig:IMAGE100-IID} and \ref{fig:IMAGE100_time}, FedHP can separately speed up training by about 2.17$\times$, 1.65$\times$, 1.06$\times$ and 2.07$\times$, compared with PENS, LD-SGD, AD-PSGD and D-PSGD.
These results demonstrate the advantage of FedHP in accelerating model training.

%which indicates the effectiveness of adaptively determining diverse and appropriate local updating frequencies and network topology for heterogeneous workers.
Secondly, we implement two sets of experiments of these algorithms on non-IID datasets.
The results of non-IID scenarios with $p$=0.6 and $p$=0.8 are presented in Fig. \ref{fig:non-IID0.6} and Fig. \ref{fig:non-IID0.8}, respectively.
We observe that FedHP can achieve the same convergence rate as that in the IID scenario while achieving higher accuracy than the other methods.
% while achieving higher accuracy, compared to the baselines.
For example, by Fig. \ref{fig:CIFAR10-non-IID0.6}, FedHP takes 5,015s to achieve 76.77\% accuracy for AlexNet on CIFAR-10, while PENS, LD-SGD, AD-PSGD and D-PSGD takes 11,953s, 8,926s, 5,539s and 10,634s to achieve 73.52\%, 70.54\%, 69.29\% and 70.35\% accuracy, respectively.
By Fig. \ref{fig:CIFAR10-non-IID0.8}, FedHP can improve the test accuracy by about 4.83\%, 13.37\%, 14.26\% and 13.52\% on CIFAR-10 with non-IID level of $p$=0.8, compared with PENS, LD-SGD, AD-PSGD and D-PSGD.
The above results indicate the effectiveness of FedHP by adaptively assigning appropriate local updating frequencies and constructing network topology for heterogeneous workers.

% By Fig. \ref{fig:CIFAR10-IID} for CNN on non-IID($p$=0.6) EMNIST, FedHP can improve the test accuracy by about 1\%, 2\%, 3\% and 4\%, compared to AD-PSGD, LD-SGD, D-PSGD and PENS, respectively, while the improvement is separately 1\%, 2\%, 3\% and 4\% for CNN on non-IID($p$=0.8) EMNIST by Fig. \ref{fig:CIFAR10-IID}.
% Besides, by Fig. \ref{fig:CIFAR10-IID} for AlexNet on non-IID($p$=0.6) Cifar10, FedHP, AD-PSGD, LD-SGD, D-PSGD, PENS separately achieve 58\% accuracy, 58\% accuracy,58\% accuracy, 58\% accuracy, 58\% accuracy, while for the non-IID(0.8) Cifar10 by Fig. \ref{fig:CIFAR10-IID}, FedHP, AD-PSGD, LD-SGD, D-PSGD, PENS achieves 58\% accuracy, 58\% accuracy,58\% accuracy, 58\% accuracy, 58\% accuracy, respectively.
% Besides, for AlexNet on non-IID($p$=0.6) Cifar10 by Fig. \ref{fig:CIFAR10-IID}, FedHP can achieve the improvement by about 6\%, 6\%, 6\% and 6\%, compared to AD-PSGD, LD-SGD, D-PSGD and PENS, respectively, while the improvement is separately 1\%, 2\%, 3\% and 4\% for CNN on non-IID($p$=0.8) EMNIST by Fig. \ref{fig:CIFAR10-IID}.

\subsubsection{Effect of Statistical Heterogeneity}
To demonstrate the robustness of FedHP to non-IID data, we show the test accuracies of these algorithms at different non-IID levels in Fig. \ref{fig:non-IID_level}, where the horizontal axis denotes the non-IID level of the datasets.
% By Fig. \ref{fig:waiting_time}, we can find that FedHP achieves much higher test accuracy than AD-PSGD, D-PSGD and LD-SGD.
% For instance, by Fig. \ref{fig:waiting_time} for non-IID($p$=0.8) IMAGE100, the model accuracy of FedHP
% is 50\% while AD-PSGD, D-PSGD and LD-SGD is 1\%, 2\% and 3\%, respectively.
By Fig.\ref{fig:non-IID_level}, we observe that the test accuracies of models trained by the five algorithms on all datasets decrease with the increasing of non-IID level.
However, FedHP can always achieve the highest model accuracy in comparison with the other algorithms.
In addition, PENS with performance-based neighbor selection can achieve higher model accuracy than the algorithms without considering the challenge of statistical heterogeneity.
For instance, by Fig. \ref{fig:IMAGE100-non-IID_level}, FedHP and PENS achieve 50.63\% and 47.81\% accuracy on IMAGE-100 with non-IID level of $p$=0.8, while LD-SGD, AD-PSGD and D-PSGD achieve 45.69\%, 45.12\% and 45.83\%, respectively.
In AD-PSGD, each worker probably receives the stale models for aggregation, which amplifies the negative impact of non-IID data on model performance, leading to the lowest test accuracy.
Both D-PSGD and LD-SGD adopt static network topologies without considering the challenge of statistical heterogeneity on model training, thus they suffer from severe loss of accuracy.
Although PENS allows workers with similar data distributions to communicate with each other in order to deal with the statistical heterogeneity, it still achieves a lower test accuracy than FedHP.
% in order to reduce the impact of non-IID data on model training
More specifically, by Fig. \ref{fig:IMAGE100-non-IID_level}, FedHP can achieve improvement of test accuracy by about 5.90\%, 10.81\%, 12.22\%, 10.47\% for VGG-16 on IMAGE-100 with non-IID level of $p$=0.8, compared with the baselines (\ie, AD-PSGD, LD-SGD, D-PSGD, PENS).
Collectively, these results demonstrate the advantage of FedHP in addressing the challenge of statistical heterogeneity.

\subsubsection{Effect of System Heterogeneity}
To further illustrate the efficiency of FedHP, the average waiting time of five algorithms on the three datasets is illustrated in Fig. \ref{fig:waiting_time},
% where the horizontal axis denotes the number of training communication rounds.
where we find that FedHP takes much less waiting time than both D-PSGD and PENS.
For instance, by Fig. \ref{fig:CIFAR10-waiting_time}, the average waiting time of FedHP is 1.7s while PENS and D-PSGD incur average waiting time of 12.1s and 10.6s, respectively.
That is because both D-PSGD and PENS assign identical local updating frequencies for workers without considering system heterogeneity, resulting in non-negligible waiting time.
In addition, PENS always suffers from more computing time for neighbor selection and model training, incurring the highest average waiting time among five algorithms.
As shown in Fig. \ref{fig:waiting_time}, the average waiting time of AD-PSGD is the lowest among these algorithms, because in the asynchronous scenario, workers update their local models as soon as they receive any models from their neighbors.
Besides, LD-SGD, implemented to alternate the frequencies of local updating and global updating, reduces the variance of waiting time to some extent.
Concretely, by Fig. \ref{fig:ImageNet-waiting_time}, FedHP and AD-PSGD only incur average waiting time of 3.2s and 2.9s, while LD-SGD, D-PSGD and PENS incur average waiting time of 19.2s, 21.5 and 24.7s, respectively.
The above results explain why FedHP and AD-PSGD can achieve much faster converge rate than D-PSGD and PENS while LD-SGD takes less completion time than D-PSGD in Figs. \ref{fig:IID}, \ref{fig:non-IID0.6} and \ref{fig:non-IID0.8}.
The results in Fig. \ref{fig:waiting_time} demonstrate that FedHP can well overcome the challenges of system heterogeneity compared with existing methods.

% \vspace{-1.2em}
\section{Related Work}\label{sec:related}
% \vspace{-0.3em}
% FL has become a practical and promising approach for distributed machine learning over distributed local data \cite{kairouz2019advances,mcmahan2017communication,li2020federated,park2019wireless,yang2019federated}.
The concept of FL was first introduced in \cite{mcmahan2017communication}, which has demonstrated the effectiveness of performing distributed model training over distributed and isolated datasets.
In order to reduce the communication resource consumption, the early works explored to optimize the local updating frequency \cite{wang2019adaptive, xu2022adaptive,li2019communication}.
% there have been some researchers studying the control of local updating frequency
% When the local update frequency is larger, the less global aggregation needs to be performed, resulting in less communication resource consumption.
As the local updating frequency increases, the frequency for global aggregation can relatively get decreased, therefore, the communication resource for model transmission can be saved to a great extent.
% And the relevant researchers on local updating frequency
However, these related researches mainly focus on PS-based FL \cite{wang2019adaptive, xu2022adaptive}, which suffers from the single point of failure problem \cite{lian2017can,yu2019parallel}.
Herein, we focus on the more attractive DFL, where Li \etal \cite{li2019communication} proposed LD-SGD to alternate the frequencies of local updating and global updating to deal with the resource-constrained issue, but they could not address the challenge of system heterogeneity.
% However, since the potential of single point failure and limited system scalability, we focuses on DFL
% However, since the single-point-of-failure problem in PS, we mainly focuses on DFL.

% The greater the local update frequency, the less global aggregation needs to be performed, resulting in less communication resource consumption.

As for network topology construction in DFL, there have been many related studies \cite{wang2019matcha,xu2021decentralized,zhou2021communication,wang2022accelerating,onoszko2021decentralized}.
Wang \etal \cite{wang2019matcha} proposed MATCHA, which uses matching decomposition sampling of the base
topology to parallelize inter-worker information exchange so as to significantly reduce communication delay.
Besides, Xu \etal \cite{xu2021decentralized} dynamically constructed an efficient P2P topology to address the challenge of resource limitation and network dynamics.
% In addition, Zhou \etal \cite{zhou2021communication} proposed NetMax, which enables workers to asynchronously communicate via high-speed links to deal with communication heterogeneity.
However, the above works all suffered from a drop in model accuracy without considering the negative effect of statistical heterogeneity.
In order to overcome statistical heterogeneity, Wang \etal \cite{wang2022accelerating} proposed CoCo to preferentially select neighbors with large differences in data distribution, while Onoszko \etal \cite{onoszko2021decentralized} proposed PENS, where workers with similar data distributions communicate with each other.
However, CoCo and PENS did not overcome the challenge of system heterogeneity, often resulting in idle time for staying and waiting for the stragglers before model aggregation.
On the contrary, FedHP investigates the benefits of controlling local updating frequency and network topology, which are jointly optimized to adequately address the issues of system and statistical heterogeneities.
\section{Conclusion}\label{sec:conclusion}
This work focuses on system heterogeneity and statistical heterogeneity for DFL.
To overcome these challenges, we have proposed FedHP to achieve fast convergence by jointly optimizing both the local updating frequency and network topology in DFL.
We have analyzed the convergence rate of FedHP and proposed an efficient algorithm.
We have evaluated the performance of FedHP through extensive simulations and the results have demonstrated the efficiency of FedHP.

\section{Acknowledgement}\label{sec:Acknowledgement}
The corresponding authors of this paper are Yang Xu.
This article is supported in part by the National Key Research and Development Program of China (Grant No. 2021YFB3301501); in part by the National Science Foundation of China (NSFC) under Grants 62102391, 62132019 and 61936015; in part by the Jiangsu Province Science Foundation for Youths (Grant No. BK20210122).

\balance
\bibliographystyle{IEEEtran}
\bibliography{content/refs}
\end{document}